\documentclass[prd,preprint,showpacs,showkeys]{revtex4-1}
\usepackage[utf8]{inputenc}
\usepackage{amsmath}
\usepackage{latexsym}
\usepackage{amsfonts}
\usepackage{graphicx}
\usepackage{mathrsfs}
\usepackage{CJK}
\usepackage{longtable}
\usepackage[normalem]{ulem}
\usepackage{slashed}

\usepackage{hyperref}
\hypersetup{
  colorlinks = true,
  urlcolor = blue,
  linkcolor = blue,
  citecolor = green,
  filecolor = magenta,
}

\newcommand{\ud}{\mathrm{d}}
\newcommand{\pd}{\partial}
\newcommand{\script}{\mathscr}
\newcommand{\order}[1]{\mathcal O\left(#1\right)}
\newcommand{\scri}{\mathscr I}
\newcommand{\tnabla}{\tilde{\nabla}}
\newcommand{\hnabla}{\bar{\nabla}}
\newcommand{\tvarphi}{\tilde{\varphi}}
\newcommand{\hvarphi}{\bar{\varphi}}
\newcommand{\tG}{\tilde{G}}
\newcommand{\tepsilon}{\tilde{\epsilon}}
\newcommand{\tg}{\tilde{g}}
\newcommand{\hg}{\bar{g}}
\newcommand{\hn}{\bar{n}}
\newcommand{\hK}{\bar{K}}
\newcommand{\hX}{\bar{X}}
\newcommand{\lie}{\mathscr L}

\begin{document}

\title{``Conserved charges'' of the Bondi-Metzner-Sachs algebra in the Brans-Dicke theory}
\author{Shaoqi Hou}
\email{hou.shaoqi@whu.edu.cn}
\affiliation{School of Physics and Technology, Wuhan University, Wuhan, Hubei 430072, China}
\author{Zong-Hong Zhu}
\email{zhuzh@whu.edu.cn (Corresponding author)}
\affiliation{School of Physics and Technology, Wuhan University, Wuhan, Hubei 430072, China}
\affiliation{Department of Astronomy, Beijing Normal University, Beijing 100875,  China}
\date{\today}

\begin{abstract}
The asymptotic symmetries in the Brans-Dicke theory are analyzed using Penrose's conformal completion method, which is independent of the coordinate system used. 
These symmetries  indeed include supertranslations and Lorentz transformations for an asymptotically flat spacetime. 
With the Wald-Zoupas formalism, ``conserved charges'' and fluxes of the Bondi-Metzner-Sachs algebra are computed.
The scalar degree of freedom contributes only to the Lorentz boost charge, even though it plays a role in various fluxes.
The flux-balance laws are further applied to constrain displacement memory, spin memory and center-of-mass memory effects. 
\end{abstract}


\maketitle

\section{Introduction}\label{sec-int}

The detection of gravitational waves (GWs) by LIGO/Virgo collaborations \cite{Abbott:2016blz,Abbott:2016nmj,Abbott:2017vtc,Abbott:2017oio,TheLIGOScientific:2017qsa,Abbott:2017gyy,LIGOScientific:2018mvr,Abbott:2020uma,LIGOScientific:2020stg,Abbott:2020khf,Abbott:2020tfl,Abbott:2020mjq,Abbott:2020niy} confirmed Einstein's prediction based on general relativity (GR) \cite{Einstein:1916cc,Einstein:1918btx}.
The GW is now a probe into the nature of gravity in the strong-field and high-speed regime. 
With the GW, there are several methods to tell whether gravity is described by GR or its alternatives.
For example, one may examine whether the GW waveform agrees with GR's prediction precisely; 
one could also count how many GW polarizations are detected \cite{Will:2014kxa,Gong:2018ybk}. 
Probably, the GW memory effect is the most intriguing phenomenon because of its intimate relation with the asymptotic symmetries.

The memory effect and the asymptotic symmetries have been studied by numerous works in GR \cite{Zeldovich:1974gvh,Braginsky:1986ia,Christodoulou1991,Bondi:1962px,Sachs:1962wk,Sachs:1962zza,Strominger:2014pwa}. 
This effect usually refers to the permanent change in the relative distance between test particles far away from the source, approximately at the null infinity $\scri$, due to the passage of GWs.
So it is also called the displacement memory.
The asymptotic symmetries are diffeomorphisms preserving the geometry of $\scri$, and form the Bondi-Metzner-Sachs (BMS) group, which is a semi-direct product of an infinite dimensional, commutative supertranslation group and the Lorentz group.
The energy flux of the GW induces a transition among degenerate vacua, which are associated with each other by the action of supertranslations.
This explains the memory effect in GR \cite{Strominger:2014pwa}.
There are also the spin memory and center-of-mass (CM) memory, which are related to the angular momentum flux arriving at $\scri$  \cite{Pasterski:2015tva,Nichols:2018qac}.

Alternative theories of gravity also possess the memory effect, as discussed in Refs.~\cite{Lang:2013fna,Lang:2014osa,Du:2016hww,Kilicarslan:2018bia,Kilicarslan:2018yxd,Kilicarslan:2018unm,Hou:2020xme}.
In particular, Ref.~\cite{Hou:2020tnd} discussed the memory effect and BMS symmetries in the Brans-Dicke theory (BD) \cite{Brans:1961sx} using the fully nonlinear equations of motion, as opposed to the post-Newtonian formalism in Refs.~\cite{Lang:2013fna,Lang:2014osa}.
It was discovered that there are also asymptotic symmetries at $\scri$ in BD, similar to those in GR.
Because of the presence of the plus and cross polarizations in BD, the displacement memory effect also exists in BD, and is related to the energy flux and supertranslations.
The breathing polarization also causes the displacement memory, and is named S memory by Du and Nishizawa \cite{Du:2016hww}.
It is due to the angular momentum flux penetrating $\scri$ and the Lorentz transformations cause the vacuum transitions in the scalar sector.
Utilizing a slightly different coordinate system, Ref.~\cite{Tahura:2020vsa} obtained similar results. 
In the current work, the asymptotic symmetries of an asymptotically flat spacetime in BD will be analyzed again using Penrose's conformal completion method \cite{Penrose:1962ij,Penrose:1965am}.
This method is covariant and independent of the coordinate system used. 

As is well-known, the existence of symmetries implies there are some conserved charges via Noether's theorem.
Thus, the BMS symmetries on $\scri$ thus prompt us to search for such quantities defined on $\scri$.
However, in general, there are GWs passing through $\scri$, so it is tricky to obtain them, and worse, none of these quantities are actually conserved.
These quantities vary along $\scri$, and the changes should be given by some fluxes.
All the ``conserved charges'' and associated fluxes can be calculated using the Hamiltonian formalism devised by Wald and Zoupas \cite{Wald:1999wa}.
This is a general method, applicable to any theory of gravity.
The procedure starts with specifying the phase space with certain boundary conditions, computing the presymplectic potential current $\theta_{abc}$ and the symplectic current $\omega_{abc}$, and obtaining the Neother charge 2-form $Q_{ab}(\xi)$ associated with an infinitesimal BMS transformation $\xi^a$.
Then, to find the ``conserved charges'' and fluxes on $\scri$, one studies the asymptotic behavior of the symplectic current so that one can construct a second presymplectic potential current $\Theta_{abc}$ on $\scri$ which gives the restriction of the symplectic current to $\scri$.
Finally, the flux density is simply $\Theta_{abc}$, and the variation of the ``conserved charge''is $\delta Q_\xi[\mathscr C]=\oint_{\mathscr C}[\delta Q_{ab}(\xi)-\xi^c\theta_{cab}+\xi^c\Theta_{cab}]$ with $\mathscr C$ a cross section of $\scri$.
Once a suitable reference spacetime is chosen, the ``conserved charge'' $Q_\xi$ can thus be obtained and satisfies $Q_\xi[\mathscr C]-Q_\xi[\mathscr C']=\int_{\mathscr B}\Theta_{abc}$ where $\mathscr B$ is a patch in $\scri$, bounded by $\mathscr C$ and $\mathscr C'$.
There are also some ambiguities in choosing $\theta_{abc}, \omega_{abc}$, and $\Theta_{abc}$, as well as the issue with choosing the reference spacetime, which are discussed in Ref.~\cite{Wald:1999wa} more carefully.
In addition, Refs.~\cite{Chandrasekaran:2018aop,Bonga:2019bim} nicely reviewed this formalism, and are worthwhile to read.

Previously, Noether charges and currents have also been considered for black holes in a more general BD with a variable $\omega(\varphi)$ and a generic potential $V(\phi)$ in both the Jordan and Einstein frames \cite{Bhattacharya:2017pqc,Bhattacharya:2018xlq}.
References~\cite{Duval:2014uva,Duval:2014lpa} found out that at least in GR, the BMS group is a subgroup of the so-called conformal Carroll group, whose charges have been computed.
One may also add to the action the terms that have no influence on the equations of motion, but that may lead to new charges as considered in Refs.~\cite{Godazgar:2020gqd,Godazgar:2020kqd}.

In this work, we apply Wald-Zoupas formalism to BD in the following sections.
We start with a brief review on the asymptotically flat spacetime in BD in Sec.~\ref{sec-bd}.
Then, the asymptotic structure is discussed again based on the conformal completion method in Sec.~\ref{sec-asy}.
There, following Refs.~\cite{Geroch1977,Ashtekar:1981hw,Ashtekar:1981bq,Ashtekar:2014zsa}, the radiative modes are identified in Sec.~\ref{sec-rad},
then, we determine the infinitesimal BMS symmetries in Sec.~\ref{sec-bms}.
Section~\ref{sec-cc-f} discusses the ``conserved charges'' and the fluxes.
The presymplectic potential current and the symplectic current are computed and analyzed in Sec.~\ref{sec-sym}.
Based on these, the fluxes and the charges can be obtained in the following two subsections~\ref{sec-flux} and \ref{sec-cc}.
In the end, the flux-balance laws are applied to constrain displacement memory (Sec.~\ref{sec-dpm}), spin memory (Sec.~\ref{sec-sm}) and CM memory (Sec.~\ref{sec-cmm}) in Sec.~\ref{sec-mem}.
Section~\ref{sec-con} is a short summary.
Some technical details have been relegated in Appendices \ref{app-fin} and \ref{app-v36}.
The abstract index notation is used \cite{Wald:1984rg}, and the speed of light $c=1$ in vacuum.


\section{Brans-Dicke theory}
\label{sec-bd}

In this section, we will review the asymptotically flat spacetime in BD based on Ref.~\cite{Hou:2020tnd}.
As is well known, the action of BD has the following form \cite{Brans:1961sx},
\begin{equation}
    \label{eq-act-bd}
    S=\frac{1}{16\pi G_0}\int\ud x^4\sqrt{-g}\left( \varphi R-\frac{\omega}{\varphi}\nabla_a\varphi\nabla^a\varphi \right),
\end{equation}
where $\omega$ is a constant, $G_0$ is the bare gravitational constant, and the matter action is ignored.
Some phenomenological aspects have been summarized in Ref.~\cite{Hou:2020tnd}.
The variational principle gives rise to the following equations of motion,
\begin{subequations}
    \label{eq-eoms}
 \begin{gather}
    R_{ab}-\frac{1}{2}g_{ab}R=\frac{8\pi G_0}{\varphi}\mathcal T_{ab},\label{eq-ein}\\ 
    \nabla_a\nabla^a\varphi=0,\label{eq-s}
\end{gather}   
\end{subequations}
in which $\mathcal T_{ab}$ is the effective stress-energy tensor for $\varphi$, given by 
\begin{equation}
    \label{eq-eff-s}
    \mathcal T_{ab}=\frac{1}{8\pi G_0}\left[\frac{\omega}{\varphi}\left(\nabla_a\varphi\nabla_b\varphi-\frac{1}{2}g_{ab}\nabla_c\varphi\nabla^c\varphi\right)+\nabla_a\nabla_b\varphi-g_{ab}\nabla_c\nabla^c\varphi\right].
\end{equation}
Equation~\eqref{eq-act-bd} is said to be written in Jordan frame.

From the previous study \cite{Hou:2020tnd}, one knows that $\varphi=\varphi_0+\order{r^{-1}}$ in an asymptotically flat spacetime. 
So one can perform the following conformal transformation $\tilde g_{ab}=\frac{\varphi}{\varphi_0}g_{ab}$, and set $\frac{\varphi}{\varphi_0}=e^{\tilde\varphi}$, then the action becomes \cite{Poisson2014}
\begin{equation}
    \label{eq-act-e}
    S=\frac{1}{16\pi\tilde G}\int\sqrt{-\tilde{g}}\left( \tilde R-\frac{2\omega+3}{2}\tnabla_a\tvarphi\tnabla^a\tvarphi \right),
\end{equation}
where $\tilde G=G_0/\varphi_0$.
This action is written in Einstein frame.
The equations of motion are given by 
\begin{subequations}
    \label{eq-eoms-e}
\begin{gather}
    \tilde R_{ab}-\frac{1}{2}\tilde g_{ab}\tilde R=8\pi G_0\tilde{\mathcal T}_{ab},\label{eq-ein-e}\\ 
    \tilde\nabla_a\tilde\nabla^a\tilde\varphi=0,\label{eq-seq-c}
\end{gather}
\end{subequations}
with 
\begin{equation}
    \label{eq-effs-c}
    \tilde{\mathcal T}_{ab}=\frac{2\omega+3}{16\pi G_0}\left( \tilde\nabla_a \tilde\varphi\tilde\nabla_b\tilde\varphi-\frac{1}{2}\tilde g_{ab}\tilde\nabla^c\tilde\varphi\tilde\nabla_c\tilde\varphi\right).
\end{equation}
In Einstein frame, $\tvarphi$ is proportional to a canonical scalar field.

As discussed in Ref.~\cite{Hou:2020tnd}, Eqs.~\eqref{eq-eoms} can be solved using the generalized Bondi-Sachs coordinates $(u,r,x^2=\theta,x^3=\phi)$ \cite{Barnich:2010eb},
\begin{equation}
    \label{eq-bc}
    \ud s^2=e^{2\beta}\frac{V}{r}\ud u^2-2e^{2\beta}\ud u\ud r+h_{AB}(\ud x^A-U^A\ud u)(\ud x^B-U^B\ud u),
\end{equation}
with $A,B=2,3$.
$\beta,V,U^A$ and $h_{AB}$ are six arbitrary functions. 
Moreover, one imposes certain boundary conditions \cite{Barnich:2010eb},
\begin{equation}
    \label{eq-exp-m}
    \beta=\order{r^{-1}},\quad V=-r+\order{r^0},\quad U^A=\order{r^{-2}},
\end{equation}
and the determinant condition,
\begin{equation}
    \label{eq-det}
    \det(h_{AB})=r^4\left(  \frac{\varphi_0}{\varphi}\right)^2\sin^2\theta.
\end{equation}
Then, one obtains the series expansions in powers of $1/r$ 
\begin{subequations}
    \label{eq-sol}
    \begin{gather}
    \varphi=\varphi_0+\frac{\varphi_1}{r}+\frac{\varphi_2}{r^2}+\order{\frac{1}{r^3}},\label{eq-exp-s}\\
        g_{uu}=-1+\frac{2m+\varphi_1/\varphi_0}{r}+\order{\frac{1}{r^2}},\\
     g_{ur}=-1+\frac{\varphi_1}{\varphi_0r}+\frac{1}{r^2}\left[ \frac{1}{16}\hat c_{A}^{B}\hat c^{A}_{B}+\frac{2\omega-5}{8}\left( \frac{\varphi_1}{\varphi_0} \right)^2+\frac{\varphi_2}{\varphi_0} \right]+\order{\frac{1}{r^3}},\\ 
     \begin{split}
         g_{uA}=&\frac{\mathscr D_B\hat c^B_A}{2}+\frac{2}{3r}\left[ N_A+\frac{1}{4}\hat c_{AB}\mathscr D_C\hat c^{BC}-\frac{\varphi_1}{12\varphi_0}\mathscr D_B\hat c^B_A \right]
         +\order{\frac{1}{r^2}},
     \end{split}\label{eq-met-uA}\\
     \begin{split}
     g_{AB}=&r^2\gamma_{AB}+r\left( \hat c_{AB}-\gamma_{AB}\frac{\varphi_1}{\varphi_0} \right)+\hat d_{AB}+\gamma_{AB}\left( \frac{1}{4}\hat c_{C}^{D}\hat c^{C}_{D}+\frac{\varphi_1^2}{\varphi_0^2}-\frac{\varphi_2}{\varphi_0} \right)
     +\order{\frac{1}{r}}.
     \end{split}
    \end{gather}
\end{subequations}
Here, $\gamma_{AB}$ is the metric on a unit 2-sphere, and $\mathscr D_A$ is its compatible covariant derivative. 
$\varphi_1,\,\varphi_2,\,\hat c_{AB}$, and $\hat d_{AB}$ are expansion coefficients, which are arbitrary functions of $(u,x^A)$. 
The indices of $\hat c_{AB}$ and $\hat d_{AB}$ are raised by $\gamma^{AB}$, and one has $\gamma^{AB}\hat c_{AB}=\gamma^{AB}\hat d_{AB}=0$.
The functions $m$ and $N_A$ of $(u,x^A)$ are called the Bondi mass aspect and the angular momentum aspect, respectively.
The Einstein's equation~\eqref{eq-ein} leads to the following  evolutions of $m$ and $N_A$,
\begin{subequations}
 \begin{equation}
    \label{eq-uu2-mdot}
    \dot {m}=-\frac{1}{4}\mathscr D_A\mathscr D_BN^{AB}-\frac{1}{8}N_{AB}N^{AB}-\frac{2\omega+3}{4}\left( \frac{N}{\varphi_0} \right)^2,
\end{equation}
\begin{equation}
    \label{eq-uA2-ndot}
    \begin{split}
    \dot {N}_A=&\mathscr D_Am
        +\frac{1}{4}(\mathscr D_B\mathscr D_A\mathscr D_C\hat c^{BC}-\mathscr D_B\mathscr D^B\mathscr D_C\hat c_A^C)\\ 
        &-\frac{1}{16}\mathscr D_A(N^{B}_{C}\hat c_{B}^{C})+\frac{1}{4}N^{B}_{C}\mathscr D_A\hat c_{B}^{C}+\frac{1}{4}\mathscr D_B(N_{A}^{C}\hat c^{B}_{C}-\hat c_{A}^{C}N^{B}_{C})\\ 
        &+\frac{2\omega+3}{8\varphi_0^2}(\varphi_1\mathscr D_AN-3N\mathscr D_A\varphi_1),
    \end{split}
\end{equation}
where $N_{AB}=-\pd\hat c_{AB}/\pd u$ is the news tensor, and $N=\pd\varphi_1/\pd u$.
Finally, the equation of motion \eqref{eq-s} for $\varphi$ gives 
\begin{equation}
    \label{eq-ev-varphis}
    \dot\varphi_2=\frac{\varphi_1N}{\varphi_0}-\frac{1}{2}\mathscr D^2\varphi_1,
\end{equation}
\end{subequations}
with $\mathscr D^2=\mathscr D_A\mathscr D^A$.

As in GR, the asymptotically flat spacetime in BD also enjoys BMS symmetries. 
An infinitesimal BMS transformation $\xi^a$ is parameterized by $\alpha(x^A)$ and $Y^A(x^B)$ defined on the unit 2-sphere. 
The transformation generated by $\alpha$ is called a supertranslation, and the one by $Y^A$ a Lorentz transformation. 
The action on the solution space can be easily computed, for instance, given by \cite{Hou:2020tnd}
\begin{subequations}
    \begin{gather}
    \delta_\xi\varphi_1=fN+\frac{\psi}{2}\varphi_1+Y^A\mathscr D_A\varphi_1,    \label{eq-bms-phi1}\\
        \delta_\xi\hat c_{AB}=-fN_{AB}-2\mathscr D_A\mathscr D_Bf+\gamma_{AB}\mathscr D^2f+\lie_Y\hat c_{AB}-\frac{\psi}{2}\hat c_{AB},\label{eq-bms-c}
    \end{gather}
and thus 
    \begin{gather}
        \delta_\xi N_{AB}=f\dot N_{AB}+\lie_YN_{AB},\\ 
        \delta_\xi N=f\dot N+\psi N+Y^A\mathscr D_AN,
    \end{gather}
where $\psi=\mathscr D_AY^A$.
\end{subequations}
With these, one can discuss the relation between BMS symmetries and gravitational memories.
It turns out that the displacement memory effect in the tensor sector is caused by the null energy fluxes, including that of the scalar field $\varphi$, passing through $\scri$, which is similar to the one in GR. 
This memory effect is associated with the supertranslation transformation, which induces the transition among the vacua in the tensor sector.
The scalar sector also contains degenerate vacua, so the displacement memory occurs in the scalar sector, too.
This is caused by the passage of the angular momentum fluxes through $\scri$, and a Lorentz transformation induces the transition among the vacua.
The spin memory and the CM memory effects are also interesting  in GR \cite{Pasterski:2015tva,Nichols:2018qac}.
Here, they exist in the tensor sector as well.
However, neither of them is present in the scalar sector.

In Ref.~\cite{Hou:2020tnd}, we did not calculate the ``conserved charges'' of the asymptotically flat spacetime in BD.
In the current work, we will compute them using the covariant phase space formulism devised by Wald and Zoupas \cite{Wald:1999wa}.
For that purpose, one starts with the asymptotic structure of BD in the next section.

\section{Asymptotic structure at null infinity}
\label{sec-asy}

The asymptotic structure of spacetimes in GR has been discussed and summarized in Refs.~\cite{Geroch1977,Ashtekar:1981bq,Wald:1984rg,Ashtekar:2014zsa}.
In this section, we will follow these treatments to study the asymptotic structure at $\scri$ in BD.
This treatment utilizes the conformal completion, which brings $\scri$ to a finite place.

The asymptotically flat spacetime at $\scri$ in BD can be defined in the following way. 
A spacetime $(M,g_{ab})$ is said to be asymptotically flat at $\scri$ in vacuum BD, if there exists an unphysical spacetime $(M',g'_{ab})$ and a conformal transformation $\mathcal C:M\rightarrow\mathcal C[M]\subset M'$ such that:
\begin{enumerate}
    \item $g'_{ab}=\Omega^2\mathcal C^*g_{ab}$ in $\mathcal C[M]$, for some conformal factor $\Omega$, where $\mathcal C^*$ is the pullback;
    \item $\scri$ is the boundary of $M$ in $M'$, and on it, $\Omega=0$ and $\nabla'_a\Omega\ne0$;
    \item the topology of $\scri$ is $\mathbb S^2\times\mathbb R$;
    \item equations \eqref{eq-eoms} are satisfied near $\scri$.
\end{enumerate}
With this definition, one can work out the asymptotic structure at $\scri$ for BD.
However, Eqs.~\eqref{eq-eoms} are very complicated due to the fact that $\varphi$ is not a canonical scalar field, so the discussion in Jordan frame would be very involved. 
Therefore, we would like to work in Einstein frame, where the equations of motion \eqref{eq-eoms-e} are simpler, and $\tilde\varphi$ is a canonical scalar field modulo a factor.
We are allowed to do the conformal completion in Einstein frame, because under the above conformal transformation relating $(M,g_{ab})$ to $(M',g'_{ab})$, one can find another unphysical spacetime $(\bar M,\bar g_{ab})$ with $\bar g_{ab}=\Omega^2\tg_{ab}$.
As a matter of fact, $\bar g_{ab}=\frac{\varphi}{\varphi_0}g'_{ab}$.
In this spacetime, $\scri$ is still the boundary of $M$ in $\bar M$ with the topology of $\mathbb S^2\times\mathbb R$, and on it, $\Omega=0$ and $\hnabla_a\Omega\ne0$, since $\hnabla_a=\nabla'_a=\pd_a$ for a scalar field.
However, instead of Eqs.~\eqref{eq-eoms}, one requires Eqs.~\eqref{eq-eoms-e} to hold near $\scri$, now.

In the following, we will first identify the radiative modes in BD, and then discuss the asymptotic symmetries.

\subsection{Radiative modes}
\label{sec-rad}

As a consequence, in Einstein frame, we will effectively perform the conformal completion for GR with a canonical scalar field.
Many results obtained in GR can be carried over directly.
For example, the conformal transformation of $\tvarphi$ is $\tvarphi=\Omega\hvarphi$ \cite{Geroch1977}.
Then, the Einstein's equation \eqref{eq-ein-e} becomes \cite{Geroch1977,Wald:1984rg}
\begin{equation}
    \label{eq-cf-ein}
    \Omega\bar S_{ab}+2\hnabla_a\hn_b-\bar f\hg_{ab}=\Omega^{-1}\bar L_{ab},
\end{equation}
where $\bar S_{ab}=\bar R_{ab}-\hg_{ab}\bar R/6$ is the Schouten tensor for $\hg_{ab}$, $\hn_a=\hnabla_a\Omega$, $\bar f=\hn_a\hn^a/\Omega$, and $\bar L_{ab}$ is given by
\begin{equation}
    \label{eq-def-l}
    \bar L_{ab}=\frac{2\omega+3}{2}\Omega^2\left(\bar{\mathcal T}_{ab}-\frac{1}{6}\hg_{ab}\bar{\mathcal T}\right),
\end{equation}
with $\bar{\mathcal T}_{ab}=\hvarphi^2\hn_a\hn_b+2\Omega\hvarphi\hn_{(a}\hnabla_{b)}\hvarphi+\Omega^2\hnabla_a\hvarphi\hnabla_b\hvarphi$, and $\bar{\mathcal T}=\hg^{ab}\bar{\mathcal T}_{ab}$.
Here and below, the index of $\hn_a$ will be raised by $\bar g^{ab}$, i.e., $\hn^a=\hg^{ab}\hn_b$.
The scalar equation \eqref{eq-seq-c} is 
\begin{equation}
    \label{eq-cf-seq}
    \Omega\hnabla_a\hnabla^a\hvarphi+\hvarphi\hnabla_a\hn^a-2\bar f\hvarphi=0.
\end{equation}
Although the right hand side of Eq.~\eqref{eq-cf-ein} carries a factor of $\Omega^{-1}$, it is vanishing on $\scri$ because $\bar L_{ab}$ vanishes faster by Eq.~\eqref{eq-def-l}.
The finiteness of Eq.~\eqref{eq-cf-ein} implies that $\hn_a\hn^a=0$, i.e., $\scri$ is null, as expected.

In addition, there is also a freedom in choosing the conformal factor.
A new conformal factor $\Omega'=\varpi\Omega$ with $\varpi>0$ is as good as the old one.
Under the this kind of gauge transformation, one can calculate that 
\begin{subequations}
\begin{gather}
    \hg'_{ab}=\varpi^2\hg_{ab},\quad \hvarphi'=\varpi^{-1}\hvarphi,\label{eq-gtrans}\\ 
    \hn'_a=\varpi\hn_a+\Omega\hnabla_a\varpi,\\ 
    \bar f'=\varpi^{-1}\bar f+2\varpi^{-2}\hn^a\hnabla_a\varpi+\varpi^{-3}\Omega(\hnabla^a\varpi)\hnabla_a\varpi.
\end{gather}
\end{subequations}
One may choose a gauge such that $\bar f'\ddot=0$, which also implies that $\hnabla'_a\hn'_b\ddot=0$ due to Eqs.~\eqref{eq-cf-ein} and \eqref{eq-def-l}.
Here, the symbol $\ddot=$ means to evaluate the equation on $\scri$.
This gauge is also called the {\it Bondi gauge}  by analogy \cite{Wald:1984rg}.
We will fix such gauge condition in the following, and drop all the prime symbols, i.e.,
\begin{equation}
    \label{eq-bondi}
    \bar f\ddot=0,\quad \hnabla_a\hn_b\ddot=0.
\end{equation}
These conditions imply that  on $\scri$, the integral curves of $\hn^a$ are the affinely parameterized null geodesics, and the null congruence is free of expansion, shear and rotation \cite{Hawking:1973uf}.
The first expression in the above equation can be rewritten as $\bar f=\Omega\vartheta$ for some function $\vartheta$ on $\bar M$, so $\hn_a\hn^a=\Omega^2\vartheta$.
The second expression in Eq.~\eqref{eq-bondi} is equivalent to
\begin{equation}
    \label{eq-nkil}
    \lie_{\hn}\hg_{ab}\ddot=0,
\end{equation}
that is, $\hn^a$ is a null Killing vector field on $\scri$.
A further gauge transformation would maintain the Bondi gauge as long as $\lie_{\hn}\varpi\ddot=0$.

From the above discussion, one knows that the structure of $\scri$ is characterized by $\hg_{ab}$ and $\hn^a$ at the ``zeroth order''.
However, these are spacetime quantities defined on $\bar M$. 
One may prefer to the intrinsic  ones to $\scri$, so let $\gamma_{ab}$ be the restriction of $\hg_{ab}$ to $\scri$.
$\hn^a$ is tangent to $\scri$, so it is naturally intrinsic to $\scri$.
Then, following the terminology of Ref.~\cite{Ashtekar:2014zsa}, the zeroth-order structure of $\scri$ is the pair $(\gamma_{ab},\hn^a)$.
This structure is universal, i.e., it is shared by any asymptotically flat spacetime at $\scri$ \cite{Geroch1977}.
Since $\gamma_{ab}\hn^b$ is the restriction of $\hg_{ab}\hn^b=\hnabla_a\Omega$ to $\scri$, $\gamma_{ab}\hn^b=0$, so $\gamma_{ab}$ is degenerate.
This is consistent with the fact that $\scri$ is null.

The first-order structure is the covariant derivative $\mathscr D_a$, induced on $\scri$ by $\hnabla_a$ \cite{Ashtekar:2014zsa}.
It satisfies 
\begin{equation}
    \label{eq-md-pro}
    \mathscr D_a\gamma_{bc}=0,\quad \mathscr D_a\hn^b=0.
\end{equation}
Some of the higher-order structures requires the following quantities that come from $\mathscr D_a$.
The curvature tensor $\mathscr R_{abc}{}^d$ can be defined for $\mathscr D_a$ in the following way.
Let $\nu_a$ be a covector field on $\scri$, then, one has 
\begin{equation}
    \label{eq-def-r}
    \mathscr D_{[a}\mathscr D_{b]}\nu_c=\frac{1}{2}\mathscr R_{abc}{}^d\nu_d.
\end{equation}
Define $\mathscr R_{abcd}=\gamma_{de}\mathscr R_{abc}{}^e$, then $\mathscr R_{ab}=\gamma^{cd}\mathscr R_{acbd}$ and $\mathscr R=\gamma^{ab}\mathscr R_{ab}$.
Here, $\gamma^{ab}$ is ``inverse'' to $\gamma_{ab}$ such that $\gamma_{ac}\gamma_{bd}\gamma^{cd}=\gamma_{ab}$.
By counting the number of the algebraically independent components of $\mathscr R_{abc}{}^d$, one may prove that there is a tensor field $\mathcal S_a{}^b$ satisfying \cite{Ashtekar:1981hw}
\begin{equation}
    \label{eq-s-pro}
    \mathcal S_b{}^a\hn^b=(\mathcal S_b{}^b-\mathscr R)\hn^a,
\end{equation}
such that 
\begin{equation}
    \label{eq-r-s}
    \mathscr R_{abc}{}^d=\gamma_{c[a}\mathcal S_{b]}{}^d+\mathcal S_{c[a}\delta_{b]}{}^d,
\end{equation}
with $\mathcal S_{ab}=\gamma_{bc}\mathcal S_{a}{}^c$.
So $\mathscr R_{abc}{}^d$ can be equivalently represented by $\mathcal S_a{}^b$.
In fact, $\mathcal S_a{}^b$ is nothing but the restriction of $\bar S_a{}^b$ to $\scri$.

Due to the topology of $\scri$, there exists a unique symmetric tensor field $\rho_{ab}$ on $\scri$ with the following properties \cite{Geroch1977},
\begin{equation}
    \label{eq-rho-pro}
    \rho_{ab}\hn^b=0,\quad \gamma^{ab}\rho_{ab}=\mathscr R, \quad \mathscr D_{[a}\rho_{b]c}=0.
\end{equation}
Now, it is ready to introduce the second-order structure, the news tensor $N_{ab}$, defined by
\begin{equation}
    \label{eq-def-news-cp}
    N_{ab}=\mathcal S_{ab}-\rho_{ab}.
\end{equation}
It is transverse $N_{ab}\hn^b=0$ and traceless $\gamma^{ab}N_{ab}=0$.
The nonvanishing of it indicates the presence of the tensor GW \cite{Hou:2020tnd}.
There also exists the scalar field $\hvarphi$ on $\scri$. 
Its Lie-drag $\bar N\equiv\lie_{\hn}\hvarphi=N/\varphi_0$ along the integral curves of $\hn^a$ signals the existence of the scalar GW penetrating $\scri$, so $\bar N$ (or, equivalently $N$) also belongs to the second-order structure of $\scri$.

Finally, the third-order structure is to be introduced.
One knows that \cite{Wald:1984rg}
\begin{equation}
    \label{eq-r-dec}
    \bar R_{abcd}=\bar C_{abcd}+\hg_{a[c}\bar S_{d]b}-\hg_{b[c}\bar S_{d]a}.
\end{equation}
In the above, we have seen the roles that $\bar S_{ab}$ plays in the asymptotic structure.
Now, consider $\bar C_{abcd}$.
Although $\bar L_{ab}$ is $\order{\Omega^2}$ near $\scri$, $\bar C_{abcd}$ still vanishes on $\scri$ according to Ref.~\cite{Geroch1977}. 
One thus introduces the following two quantities,
\begin{equation}
    \label{eq-def-ks}
    K^{ab}=-4\Omega^{-1}\bar C^{acbd}\hn_c\hn_d,\quad {}^*K^{ab}=-4\Omega^{-1}{}^*\bar C^{acbd}\hn_c\hn_d,
\end{equation}
where ${}^*\bar C^{acbd}$ is the Hodge dual \cite{Wald:1984rg}.
Since $K^{ab}\hn_b={}^*K^{ab}\hn_b=0$, they are naturally intrinsic to $\scri$.
They are symmetric and traceless $\gamma_{ab}K^{ab}=\gamma_{ab}{}^*K^{ab}=0$.
They are also dual to each other in the following sense,
\begin{equation}
    \label{eq-dual-ks}
    \gamma_{ac}K^{cb}=-\bar\epsilon_{acd}\hn^d{}^*K^{cb},\quad \gamma_{ac}{}^*K^{cb}=\bar\epsilon_{acd}\hn^dK^{cb},
\end{equation}
where $\bar\epsilon_{abc}$ is the volume element  on $\scri$, induced from $\bar\epsilon_{abcd}(=4\bar\epsilon_{[abc}\hn_{d]})$.
Following the argument in Ref.~\cite{Geroch1977}, it can be shown that 
\begin{subequations}
    \label{eq-pro-ks}
    \begin{gather}
    \mathscr D_{[a}\mathcal S_{b]}{}^c=\frac{1}{4}\bar\epsilon_{abd}{}^*K^{dc},\\ 
     \mathscr D_bK^{ab}=\frac{2(2\omega+3)}{3}\left[\hvarphi\lie_{\hn}\bar N-2\bar N^2\right]\hn^a,\\
       \mathscr D_b{}^*K^{ab}=0. 
    \end{gather}
\end{subequations}
${}^*K^{ab}$ is the third-order structure on $\scri$.

The gauge transformations of the above structures are given by \cite{Ashtekar:1981bq}
\begin{subequations}
    \begin{gather}
        \gamma'_{ab}=\varpi^2\gamma_{ab},\quad \hn'^a=\varpi^{-1}\hn^a,\label{eq-gt-qn}\\ 
        \mathscr D'_a\nu_b=\mathscr D_a\nu_b-2\varpi^{-1}\nu_{(a}\mathscr D_{b)}\varpi+\varpi^{-1}\gamma_{ab}\varpi^c\nu_c,\label{eq-gt-cov}\\ 
        N'_{ab}=N_{ab},\quad\bar N'=\varpi^{-2}\bar N,\label{eq-gt-news}\\
        K'^{ab}=\varpi^{-5}K^{ab},\quad {}^*K'^{ab}=\varpi^{-5}{}^*K^{ab},
    \end{gather}
\end{subequations}
where $\varpi^a$ is the restriction of $\hnabla^a\varpi$ to $\scri$.
Now, consider a special gauge transformation with $\varpi=1$ on $\scri$.
Then, the first-order structure $\mathscr D_a$ changes according to, 
\begin{equation}
    \label{eq-gt-cov-1}
    \mathscr D'_a\nu_b=\mathscr D_a\nu_b+\kappa \gamma_{ab}\hn^c\nu_c,
\end{equation}
where $\hnabla^a\varpi\ddot=\kappa \hn^a$ for some function $\kappa$ on $\scri$, but the remaining structures stay the same.
Therefore, although the zeroth-order structure does not change, i.e., $(\gamma'_{ab},\hn'^a)=(\gamma_{ab},\hn^a)$, the covariant derivatives $\mathscr D'_a$ and $\mathscr D_a$ can be different. 
This suggests to introduce the concept of an equivalence class $\{\mathscr D_a\}$, which is the set of covariant derivatives associated with each other via Eq.~\eqref{eq-gt-cov-1} \cite{Ashtekar:1981hw}.
The radiative degrees of freedom are encoded in $\{\mathscr D_a\}$.
Now, let $\hg_{ab}$ and $\hg'_{ab}$ be two metric fields in the unphysical spacetime $\bar M$, and their covariant derivatives are $\hnabla_a$ and $\hnabla'_a$, respectively.
Let  $\{\mathscr D_a\}$ and $\{\mathscr D'_a\}$ be two equivalence classes of the induced covariant derivatives from $\hnabla_a$ and $\hnabla'_a$, respectively.
Their difference is completely characterized by a symmetric tensor field $\sigma_{ab}$ with $\sigma_{ab}\hn^b=0$ and $\gamma^{ab}\sigma_{ab}=0$. 
If one introduces a covector field $\ell_a$ on $\scri$ such that $\hn^a\ell_a=1$, one can show that $\sigma_{ab}$ is the traceless part of the following tensor \cite{Ashtekar:1981hw},
\begin{equation}
    \label{eq-def-Sig}
    \Sigma_{ab}=(\mathscr D'_a-\mathscr D_a)\ell_b,
\end{equation}
where $\mathscr D'_a$ and $\mathscr D_a$ are two representatives of $\{\mathscr D'_a\}$ and $\{\mathscr D_a\}$, respectively.
One can easily verify that $\sigma_{ab}$ has two independent components, and they represent the radiative degrees of freedom in the tensor sector. 
In fact, by replacing $\nu_b$ with $\ell_b$ and substituting Eq.~\eqref{eq-r-s} in Eq.~\eqref{eq-def-r}, and contracting both sides of the result by $\hn^b$, one finds out that 
\begin{equation}
    \label{eq-lns-news}
    N_{ab}=-2\lie_{\hn}\sigma_{ab}.
\end{equation} 
Here, in order to derive this relation, one makes use of a trivial derivative $\mathring{\mathscr D}_a$ with $\mathring{\mathscr D}_a\ell_b=0$, and sets $\Sigma_{ab}=(\mathscr D_a-\mathring{\mathscr D}_a)\ell_b=\mathscr D_a\ell_b$.
In this sense, $\sigma_{ab}$ is the shear of an null congruence with tangent vector fields $\ell^a=\hg^{ab}\ell_b$ on $\scri$.

The metric solution exhibited in the previous section is actually in the Bondi gauge. 
To show this, one first transforms the solution to the one in Einstein frame, and then, performs a conformal transformation with $\Omega=1/r$. 
In the coordinates $\{u,\Omega,\theta,\phi\}$, the metric is,
\begin{equation}
    \label{eq-cf-bs}
    \begin{split}
    \ud \bar s^2=&[-\Omega^2+2\Omega^3m+\order{\Omega^4}]\ud u^2+2[1+\order{\Omega^2}]\ud u\ud\Omega+\left[ \Omega^2\mathscr D_B\hat c_A^B+\order{\Omega^3} \right]\ud u\ud x^A\\
    &+\left[\gamma_{AB}+\Omega\hat c_{AB}+\Omega^2\left( \hat d_{AB}+\frac{\varphi_1}{\varphi_0}\hat c_{AB}+\frac{\gamma_{AB}}{4}\hat c_C^D\hat c^C_D \right)+\order{\Omega^3}\right]\ud x^A\ud x^B,
    \end{split}
\end{equation}
and the scalar field is 
\begin{equation}
    \label{eq-cf-sc}
    \bar\varphi=\frac{\varphi_1}{\varphi_0}+\Omega\left( \frac{\varphi_2}{\varphi_0}-\frac{\varphi_1^2}{2\varphi_0^2}\right)+\order{\Omega^2}.
\end{equation}
With Eq.~\eqref{eq-cf-bs}, one can verify the validity of the Bondi gauge condition \eqref{eq-bondi}.
One also knows that $\hn_a=\hnabla_a\Omega$, so $\hn^a=(\pd_u)^a$.
And finally, by setting $\ell_a=(\ud u)_a+\order{\Omega}$, one gets $\sigma_{AB}=\hat c_{AB}/2$, and $N_{AB}=-\pd_u\hat c_{AB}$.

\subsection{BMS generators}
\label{sec-bms}

As discussed in Ref.~\cite{Geroch:1981ut}, an infinitesimal asymptotic symmetry $\xi^a$ induces the following variation in $\tg_{ab}$,
\begin{equation}
    \label{eq-def-bms}
    \Omega^2\delta_\xi\tg_{ab}=\Omega^2\lie_{\xi}\tg_{ab}=\lie_\xi\hg_{ab}-2\hK\hg_{ab}=2\Omega \hX_{ab},
\end{equation}
for some smooth scalar field $\hK=\xi^a\hn_a/\Omega$ \footnote{Note that $\hK=\psi/2$ defined in \cite{Hou:2020tnd}.} and some smooth tensor field $\hX_{ab}$ in $\bar M$. 
The well-posedness of this expression requires that $\xi^a\hn_a\ddot=0$, so $\xi^a$ is tangent to $\scri$.
This equation can be rewritten as
\begin{equation}
    \label{eq-xi-hg}
    \lie_\xi\hg_{ab}=2(\hK\hg_{ab}+\Omega\hX_{ab}).
\end{equation}
By examining $(\lie_\xi\lie_{\hn}-\lie_{\hn}\lie_\xi-\lie_{[\xi,\hn]})\hg_{ab}=0$ with the conformal Einstein's equation \eqref{eq-cf-ein}, one obtains that 
\begin{equation}
    \label{eq-glinkage-3}
 -\hnabla_a\hnabla_b\hK+4\hn_{(a}\hX_{b)}+2\Omega\hnabla_{(a}\hX_{b)}-\hg_{ab}\hn_c\hX^c-\frac{1}{2}\lie_\xi( \bar S_{ab}-\Omega^{-2}\bar L_{ab})-\lie_{\hn} \hX_{ab}=0,
\end{equation}
where $\hX_a=\Omega^{-1}\hX_{ab}\hn^b$ and $\hX=\hg^{ab}\hX_{ab}$.
Again, the well-posedness of Eq.~\eqref{eq-glinkage-3} leads to the fact that $\hX_{ab}$ is transverse to $\hn^a$ so that $\hX_a$ is finite on $\scri$.
The  action of $\xi^a$ on $\hn^a$ can be easily calculated, which is 
\begin{equation}
    \label{eq-def-xin}
    \lie_\xi\hn^a=-\hK\hn^a+\Omega\hnabla^a\hK-2\Omega^2\hX^a.
\end{equation}
Contracting both sides by $\hn_a$ gives 
\begin{equation}
    \label{eq-ln-k}
    \lie_{\hn}\hK=\frac{1}{2}(\lie_\xi\bar f-\hK\bar f),
\end{equation}
without imposing the Bondi gauge condition.
What about the action of $\xi^a$ on $\hvarphi$? 
First, one can perform $\delta_\xi\tvarphi=\lie_\xi\tvarphi=\Omega(\lie_\xi\hvarphi+\hK\hvarphi)$.
Second, according to the definition of the asymptotic symmetry in Ref.~\cite{Hou:2020tnd}, the transformed ``physical'' $\tvarphi$ is allowed to decay as $1/r\sim\Omega$.
Therefore, one knows that 
\begin{equation}
    \label{eq-def-vp}
    \delta_\xi\hvarphi=\lie_\xi\hvarphi+\hK\hvarphi,
\end{equation}
which actually agrees with the transformation property of $\varphi_1$ in Ref.~\cite{Hou:2020tnd}.
Indeed, $\hvarphi\ddot=\varphi_1/\varphi_0$.

Now, one knows how a BMS generator acts on $\hg_{ab}$ and $\hn^a$ in the unphysical spacetime $\bar M$ according to Eqs.~\eqref{eq-xi-hg} and \eqref{eq-def-xin}.
By restricting these equations to $\scri$, one obtains \cite{Geroch1977}
\begin{equation}
    \label{eq-xi-hg-hn}
    \lie_\xi \gamma_{ab}=2\bar K\gamma_{ab},\quad \lie_\xi\hn^a=-\bar K\hn^a.
\end{equation}
And Eq.~\eqref{eq-ln-k} implies that $\lie_\xi\bar K=0$ on $\scri$ in Bondi gauge.
Therefore, $\xi^a$ is a conformal Killing vector field on $\scri$.
As is known, among the BMS generators, there are infinitesimal supertranslations, which are given by \cite{Geroch:1981ut}
\begin{equation}
    \label{eq-st-exp}
    \xi^a=\alpha\hn^a-\Omega\hnabla^a\alpha+\Omega^2u^a,
\end{equation}
where $\alpha$ is a smooth function and $u^a$ is a smooth vector field on $\bar M$.
Moreover, $\alpha$ should satisfy $\lie_{\hn}\alpha=\Omega\varsigma_\alpha$ for some smooth function $\varsigma_\alpha$ on $\bar M$.
One can show that 
\begin{subequations}
    \begin{gather}
        \hK=\Omega(\alpha\vartheta-\varsigma_\alpha+\varrho),\\ 
  \hX_{ab}=-\bar\nabla_a\bar\nabla_b\alpha-\frac{1}{2}(\alpha\vartheta-2\varsigma_\alpha+2\varrho)\bar g_{ab}-\frac{\alpha}{2} (\bar S_{ab}-\Omega^{-2}\bar L_{ab})+2\hn_{(a}u_{b)}+\Omega\hnabla_{(a}u_{b)},\label{eq-hXab}\\
  \begin{split}
  \hX_a=&\frac{1}{2}\bar\nabla_a(\alpha\vartheta-2\varsigma_\alpha+\varrho)-\frac{1}{2}(\bar S_{ab}-\Omega^{-2}\bar L_{ab})\bar\nabla^b\alpha+\frac{1}{2}\hn^b\hnabla_bu_a\\&+\frac{\Omega}{4}[3\vartheta u_a+(\bar S_{ab}-\Omega^{-2}\bar L_{ab})u^b],
  \end{split}\\
  \hX=-\bar\nabla^2\alpha-2(\alpha\vartheta-2\varsigma_\alpha+\varrho)-\alpha\left(\frac{\bar R}{3}-\Omega^{-2}\bar L  \right)+\Omega\hnabla_au^a,
    \end{gather}
\end{subequations}
where $\varrho=u^a\hn_a$ and $u_a=\hg_{ab}u^b$.
On $\scri$, one has $\xi^a\ddot=\alpha\hn^a$ and $\lie_{\hn}\alpha\ddot=0$.
Since $\bar K\ddot=0$, $\alpha\hn^a$ is a Killing vector field.

For a generic BMS generator $\xi^a$, let us directly consider its restriction to $\scri$.
One knows that it satisfies the following conditions \cite{Geroch1977}
\begin{equation}
    \label{eq-bms-cond}
    \hn^a\xi_a=0,\quad \mathscr D_{(a}\xi_{b)}=\hK \gamma_{ab}, \quad\lie_{\hn}\xi_a=0,
\end{equation}
with $\xi_a=\hg_{ab}\xi^b$.
The first expression is because $\xi^a$ is tangent to $\scri$. 
The second and the third are basically Eqs.~\eqref{eq-xi-hg-hn}.
Conversely, if a covector field $\xi_a$ satisfies Eqs.~\eqref{eq-bms-cond}, one can find a BMS generator $\xi^a$ satisfying Eqs.~\eqref{eq-xi-hg-hn} and $\xi_a=\gamma_{ab}\xi^b$. 
Due to the degeneracy of $\gamma_{ab}$, $\xi^a$ is not unique: one can add to it an arbitrary supertranslation generator $\alpha\hn^a$ without modifying $\xi_a$.
If $\xi^a$ and $\xi'^a$ are said to be equivalent as long as they differ by a supertranslation, the solutions to Eqs.~\eqref{eq-bms-cond} belong to an equivalence class. 
The set of such equivalence classes is isomorphic to the Lorentz algebra, due to the topology of $\scri$.
Since this set is also the quotient algebra of the BMS algebra modulo the supertranslation algebra, one verifies that the BMS algebra is indeed the semi-direct sum of the supertranslation algebra and the Lorentz algebra.

Once a foliation of $\scri$ is chosen, $\xi^a$ can be uniquely decomposed.
This foliation can be obtained by starting with a reference leaf $\mathscr C_0$, a cross section, at some retarded time $u_0$, then Lie-dragging it along the integral curves of $\hn^a$ to an arbitrary $\mathscr C$.
One can further let the normal to $\mathscr C$ be $\ell_a$, then $\xi^a$ is decomposed according to,
\begin{equation}
    \label{eq-xi-dec}
    \xi^a\ddot=\left( \alpha+\frac{u}{2}\mathscr D\cdot Y \right)\hn^a+Y^a,
\end{equation}
where $\mathscr D\cdot Y=\gamma^{ab}\mathscr D_aY_b=2\hK$ on $\scri$.
Here, the component $Y^a$ is tangent to $\mathscr C$, generating the infinitesimal Lorentz transformation and leaving $\mathscr C$ invariant, but $\alpha\hn^a$, an infinitesimal supertranslation, induces a one-parameter group of diffeomorphisms that changes the foliation for a general $\alpha$.
One can check that $\lie_Y\gamma_{ab}=2\bar K\gamma_{ab}$ but $\lie_Y\hn^a=0$, so $Y^a$ itself is not a BMS generator. 
This explains the presence of the term proportional to $u$, which, together with $Y^a$, is a genuine BMS generator.

One should also know how a BMS generator transforms $\mathscr D_a$ in order to calculate the flux and the ``conserved charge''.
For this end, one  first finds out that for any $\xi^a$ and $\nu_a$, 
\begin{equation}
    \label{eq-vd-gen}
    (\lie_\xi\mathscr D_a-\mathscr D_a\lie_\xi)\nu_b=(\xi^d\mathscr R_{dab}{}^c-\mathscr D_a\mathscr D_b\xi^c)\nu_c.
\end{equation}
So for a supertranslation $\xi^a\ddot=\alpha\hn^a$, one has the following useful result,
\begin{equation}
    \label{eq-var-sig-2}
    \delta_{\alpha\hn}\Sigma_{ab}=(\lie_{\alpha\hn}\mathscr D_a-\mathscr D_a\lie_{\alpha\hn})\ell_b
        =-\mathscr D_a\mathscr D_b\alpha-\frac{\alpha}{2}N_{ab}+\kappa'\gamma_{ab},
\end{equation}
where Eq.~\eqref{eq-r-s} has been used, and $\kappa'$ is some function on $\scri$ and irrelevant for the coming discussion.
Then, the Lorentz transformation also transforms $\mathscr D_a$, which is given by \cite{Alessio:2019cae}
\begin{equation}
    \label{eq-var-sig-lt}
  \begin{split}
  \delta_Y\Sigma_{ab}=(\lie_\xi\mathscr D_a-\mathscr D_a\lie_\xi)\ell_b
  =&-\frac{u}{2}\mathscr D_a\mathscr D_b(\mathscr D\cdot Y)-\frac{1}{2}\sigma_{ab}\mathscr D\cdot Y+\lie_Y\sigma_{ab}-\frac{u}{4}(\mathscr D\cdot Y)N_{ab}\\ 
  &-\ell_{(a}\mathscr D_{b)}(\mathscr D\cdot Y)+\frac{1}{4}\gamma_{ab}\ell\mathscr D\cdot Y+\frac{1}{2}\gamma_{ab}Y^c\mathscr D_c\ell,
  \end{split}
\end{equation}
where $\ell=\gamma^{ab}\mathscr D_a\ell_b$.
The traceless parts of Eqs.~\eqref{eq-var-sig-2} and \eqref{eq-var-sig-lt} are $\delta_{\alpha\hn}\sigma_{ab}$ and $\delta_Y\sigma_{ab}$, respectively.
In the end, one knows that $\xi^a$ induces the variation of $\hg_{ab}$ according to Eq.~\eqref{eq-xi-hg}, so its connection $\bar\Gamma^c{}_{ab}$ also changes, given by
\begin{equation}
    \label{eq-var-cov-st}
    \begin{split}
        (\lie_\xi\hnabla_a-\hnabla_a\lie_\xi)\nu_b
        =&-\nu_c\delta_\xi\bar\Gamma^c{}_{ab}\\ 
        =&\nu_c[\hn^c\hX_{ab}-2\hn_{(a}\hX_{b)}{}^c-2\delta^c_{(a}\hnabla_{b)}\hK+\hg_{ab}\hnabla^c\hK+\order{\Omega}].
    \end{split}
\end{equation}
Now, take the restriction to $\scri$ and set $\nu_a=\ell_a$ to lead to 
\begin{equation}
    \label{eq-x-vd}
     \hX_{ab}=(\lie_\xi\mathscr D_a-\mathscr D_a\lie_\xi)\ell_b+2\ell_{(a}\mathscr D_{b)}\hK-\gamma_{ab}\ell_c\hnabla^c\hK,
\end{equation}
where $\hnabla^c$ is not replaced by $\mathscr D_a$ in the last term, because this term is useless in the following calculation.

\section{``Conserved charges'' and fluxes}
\label{sec-cc-f}


\subsection{(Pre)symplectic currents}
\label{sec-sym}

Following Ref.~\cite{Wald:1999wa}, one starts with the variation of the action \eqref{eq-act-bd},
\begin{equation}
    \delta S=\frac{1}{16\pi G_0}\int\ud^4\sqrt{-g}(E_{ab}\delta g^{ab}+E_\varphi\delta\varphi)+\int\ud^4x\sqrt{-g}\nabla_a\theta^a,
\end{equation}
where $E_{ab}$ is the Einstein's equation, taking a different form than but equivalent to Eq.~\eqref{eq-ein}, and $E_\varphi=R+\frac{2\omega}{\varphi}\nabla_a\nabla^a\varphi-\frac{\omega}{\varphi^2}\nabla_a\varphi\nabla^a\varphi$.
The last term above is a surface term, where $\theta^a$, or its Hodge dual is  the so-called presymplectic potential current, given by
\begin{equation}
    \label{eq-presym}
    \begin{split}
    \theta_{abc}(\delta g,\delta\varphi)=\frac{1}{16\pi G_0}\epsilon_{dabc}\bigg[&\varphi g^{de}g^{fh}(\nabla_f\delta g_{eh}-\nabla_e\delta g_{fh})\\
    & +g^{de}g^{fh}(\delta g_{fh}\nabla_e\varphi-\delta g_{eh}\nabla_f\varphi)-\frac{2\omega}{\varphi}\delta \varphi\nabla^d\varphi\bigg].
    \end{split}
\end{equation}
With $\theta_{abc}$, the symplectic current is given by,
\begin{equation}
    \label{eq-sym}
    \begin{split}
        \omega_{abc}=&\delta\theta_{abc}(\delta' g,\delta'\varphi)-\delta'\theta_{abc}(\delta g,\delta\varphi)\\ 
        =&\frac{1}{16\pi G_0}\epsilon_{dabc}\varphi w^d+\frac{1}{16\pi G_0}\epsilon_{dabc}\bigg[2g^{d[e}g^{f]h}(\delta\varphi\nabla_f\delta' g_{eh}-\delta'g_{eh}\nabla_f\delta\varphi)\\ 
        &+(g^{dp}g^{eq}g^{fh}+g^{de}g^{fp}g^{qh})\delta g_{pq}\delta'g_{eh}\nabla_{f}\varphi+\frac{1}{2}g^{fh}\delta'g_{fh}\delta g^{de}\nabla_e\varphi\\
        &-\frac{2\omega}{\varphi}\delta'\varphi\left(\delta g^{de}\nabla_e\varphi+\nabla^d\delta\varphi+\frac{1}{2}g^{ef}\delta g_{ef}\delta'\varphi\nabla^d\varphi\right)-\langle\delta\leftrightarrow\delta'\rangle\bigg],
    \end{split}
\end{equation}
where $\langle\delta\leftrightarrow\delta'\rangle$ represents the terms obtained by switching $\delta$ and $\delta'$ of the remaining terms in the square brackets, and $w^a$ has been calculated in Ref.~\cite{Wald:1999wa} for GR, i.e,
\begin{equation}
    \label{eq-wgr}
    w^a=(g^{a[e}g^{d]c}g^{bf}+g^{ae}g^{b[f}g^{c]d}+g^{a[d}g^{b]c}g^{ef})(\delta'g_{bc}\nabla_d\delta g_{ef}-\delta g_{bc}\nabla_d\delta'g_{ef}).
\end{equation}
However, the above results were computed in Jordan frame, where the Eqs.~\eqref{eq-eoms} are complicated and the calculation of the ``conserved charges'' and fluxes is likely also very involved. 

To resolve the complication, one would like to replace all quantities in Eqs.~\eqref{eq-presym} and \eqref{eq-sym} by the corresponding ones in Einstein frame. 
One may also directly calculate the presymplectic potential current and the symplectic current using the action~\eqref{eq-act-e} in Einstein frame, which are
\begin{subequations}
    \label{eq-th-ow}
    \begin{gather}
    \tilde\theta_{abc}=\frac{1}{16\pi\tilde G}\tepsilon_{dabc}[\tg^{de}\tg^{fh}(\tnabla_f\delta\tg_{eh}-\tnabla_e\delta\tg_{fh})-(2\omega+3)\delta\tvarphi\tnabla^d\tvarphi],\label{eq-th}\\
     \begin{split}
    \tilde\omega_{abc}=&\frac{1}{16\pi\tG}\tepsilon_{dabc}\tilde w^d\\ 
    &-\frac{2\omega+3}{16\pi\tG}\tepsilon_{dabc}\left[ \delta'\tvarphi\tnabla^d\delta\tvarphi+\delta'\tvarphi\delta \tg^{de}\tnabla_e\tvarphi+\frac{1}{2}\tg^{ef}\delta \tg_{ef} \delta'\tvarphi\tnabla^d\tvarphi-\langle\delta\leftrightarrow\delta'\rangle\right],\label{eq-ow}
    \end{split}   
    \end{gather}
\end{subequations}
where $\tilde w^a$ takes the similar forms to $w^a$ in Eq.~\eqref{eq-wgr} with all $g$'s and $\nabla$'s replaced by $\tg$ and $\tnabla$, respectively.
However, careful examination shows that these two methods give distinct presymplectic potential currents,
\begin{subequations}
    \begin{gather}
        \theta_{abc}(\delta g,\delta\varphi)=\tilde\theta_{abc}(\delta\tg,\delta\tvarphi)+\Delta_{abc},\\
        \Delta_{abc}\equiv\frac{3}{16\pi\tG}\tepsilon_{dabc}\left( \delta\tilde g^{de}\tnabla_e\tvarphi+\frac{1}{2}\tg^{ef}\delta\tilde g_{ef}\tnabla^d\tvarphi+\tnabla^d\delta\tvarphi \right).
    \end{gather}
\end{subequations}
Nevertheless, the symplectic currents are the same, i.e., $\omega_{abc}(\delta g,\delta\varphi)=\tilde\omega_{abc}(\delta\tg,\delta\tvarphi)$.
Although $\Delta_{abc}$ is nonvanishing in general, it is closed, $\tnabla_{[a}\Delta_{bcd]}=0$, if Eqs.~\eqref{eq-eoms-e} and their linear perturbations are satisfied.
Indeed, one can find out that 
\begin{equation}
    \label{eq-0ons}
    \begin{split}
    \tepsilon^{abcd}\tnabla_a\Delta_{bcd}=&\frac{9}{8\pi\tG}\left[ \frac{1}{2}\tg^{ab}\delta\tg_{ab}\tnabla^2\tvarphi\right.\\
    &\left.+\left( \delta\tg^{ab}\tnabla_a\tnabla_b\tvarphi+\tnabla_a\delta\tg^{ab}\tnabla_b\tvarphi+\frac{1}{2}\tg^{ab}\tnabla_c\delta\tg_{ab}\tnabla^c\tvarphi+\tnabla^2\delta\tvarphi \right) \right],
    \end{split}
\end{equation}
and the second line is the linearized scalar field equation.
Therefore, $\Delta_{abc}=3\tnabla_{[a}\mathcal Y_{bc]}$ locally for some 2-form $\mathcal Y_{ab}$, locally constructed out of $\tg_{ab},\tvarphi$ and their variations \cite{Wald:1990closed}.
According to Ref.~\cite{Wald:1999wa}, there is always an ambiguity in choosing $\theta_{abc}$. 
Since we will work in the Einstein frame, we ignore the difference $\Delta_{abc}$.

Now, choose an arbitrary, closed, embedded 3-dimensional hypersurface $\Sigma$ without boundary. 
The presymplectic form $\Xi_\Sigma$ is given by the following integral,
\begin{equation}
    \label{eq-def-presym}
    \Xi_\Sigma(\delta\tg,\delta\tvarphi;\delta'\tg,\delta'\tvarphi)=\int_\Sigma\tilde\omega_{abc}(\delta\tg,\delta\tvarphi;\delta'\tg,\delta'\tvarphi).
\end{equation} 
Suppose $\delta'\tg_{ab}$ and $\delta'\tvarphi$ is induced by a vector field $\xi^a$, that is, $\delta'\tg_{ab}=\lie_\xi\tg_{ab}$ and $\delta'\tvarphi=\lie_\xi\tvarphi$. 
Furthermore, if the equations of motion \eqref{eq-eoms-e} are satisfied by $\tg_{ab}$ and $\tvarphi$, and the linearized equations of motion are also satisfied by $\delta\tg_{ab}$ and $\delta\tvarphi$, then the above integral defines the variation of a Hamiltonian, or a charge $Q_\xi$, conjugate to $\xi^a$,
\begin{equation}
    \label{eq-def-dh}
    \slashed\delta Q_\xi[\Sigma]=\int_\Sigma\tilde\omega_{abc}(\delta\tg,\delta\tvarphi;\lie_\xi\tg,\lie_\xi\tvarphi).
\end{equation}
It turns out that the integral above can be rewritten as the one over a 2-dimensional surface $\pd\Sigma$ \cite{Wald:1999wa},
\begin{equation}
    \label{eq-dh-sur}
    \slashed\delta Q_\xi[\pd\Sigma]=\int_{\pd\Sigma}[\delta\tilde Q_{ab}-\xi^c\tilde\theta_{cab}(\delta\tg,\delta\tvarphi)],
\end{equation}
so now, we take $\slashed\delta Q_\xi$ as a function of $\pd\Sigma$, instead of $\Sigma$.
In the above expression, the Noether charge 2-form $\tilde Q_{ab}$ is 
\begin{equation}
    \label{eq-noec-2}
    \tilde Q_{ab}=-\frac{1}{16\pi\tG}\tepsilon_{abcd}\tnabla^c\xi^d,
\end{equation}
which takes exactly the same form as in GR \cite{Iyer:1994ys}.
Here, the symbol $\slashed\delta$ means that a function $Q_\xi$ might not exist.
The sufficient and necessary condition for the existence of $Q_\xi$ on $\Sigma$ is that for all $(\delta \tg_{ab},\delta\tvarphi)$ and $(\delta'\tg_{ab},\delta'\tvarphi)$ satisfying the linearized equations of motion \cite{Wald:1999wa}, 
\begin{equation}
    \label{eq-wz-18}
    \int_{\pd\Sigma}\xi^c\tilde\omega_{cab}(\delta\tg,\delta\tvarphi;\delta'\tg,\delta'\tvarphi)=0.
\end{equation}
When this condition is violated, for example, when $\pd\Sigma$ is a cross section of $\scri$, there is a prescription to find a ``conserved charge'' $Q_\xi$ conjugate to $\xi^a$ to be discussed in the next two subsections.
Before that, one has to analyze the behaviors of $\tilde\theta_{abc},\,\tilde\omega_{abc}$, and $\tilde Q_{ab}$ near $\scri$.

Equations~\eqref{eq-th-ow} and \eqref{eq-noec-2} are the most important for calculating the ``conserved charges'' at $\scri$.
Since $\scri$ in the physical spacetime is not at a finite place, it is probable that these equations blow up at $\scri$. 
So one needs to check whether they are finite at $\scri$ or not.
One also needs to know the behaviors of $\delta\tg_{ab}$ and $\delta\tvarphi$.
For that purpose, one should realize that the field variation should not change the conformal factor, $\delta\Omega=0$.
At the same time, $\scri$ is a universal structure for any asymptotically flat spacetime \cite{Geroch1977}.
So one requires that the unphysical metric $\hg_{ab}$ remain the same at $\scri$,
\begin{equation}
    \label{eq-vhg0}
    \delta\hg_{ab}=\Omega^2\delta\tg_{ab}\ddot=0,
\end{equation}
which implies that there exists a smooth tensor field $\tau_{ab}$ such that 
\begin{equation}
    \label{eq-set-vhg}
    \delta\hg_{ab}=\Omega\tau_{ab},\quad\delta\tg_{ab}=\Omega^{-1}\tau_{ab}.
\end{equation}
By the similar method to obtain Eq.~\eqref{eq-x-vd}, one can show that $\tau_{ab}\ddot=2\delta\Sigma_{ab}$.
As discussed in Sec.~\ref{sec-asy}, the Bondi gauge condition \eqref{eq-bondi} is used for simplicity. 
This condition should be  preserved under the field variation $\delta\hg_{ab}$, so one finds out that there exists a smooth covector field $\tau_a$, such that 
\begin{equation}
    \label{eq-ttp}
    \tau_{ab}\hn^b=\Omega\tau_a.
\end{equation}
Finally, there are no requirements on $\delta\tvarphi$, so one simply writes $\delta\tvarphi=\Omega\delta\hvarphi$.

Now, it is straightforward to reexpress Eqs.~\eqref{eq-th-ow} and \eqref{eq-noec-2}  in the unphysical spacetime $(\bar M,\hg_{ab})$.
Firstly, the presymplectic potential current is,
\begin{equation}
    \label{eq-th-cf}
    \tilde\theta_{abc}=\frac{1}{16\pi\tilde G_0}\bar\epsilon_{abcd}\left\{ \Omega^{-1}\left[ \hnabla_e\tau^{de}-\hnabla^d\tau-3\tau^d-(2\omega+3)\chi\hvarphi\hn^d \right]-(2\omega+3)\chi\hnabla^d\hvarphi\right\},
\end{equation}
where $\tau=\hg^{ab}\tau_{ab}$, and $\chi=\delta\hvarphi$.
Though formally, this expression blows up at $\scri$ due to $\Omega^{-1}$ factor inside the curly brackets,  it actually does not.
To show this, one starts with Einstein's equation \eqref{eq-cf-ein} in the unphysical spacetime without imposing the Bondi gauge explicitly, and then varies it,
\begin{equation}
    \label{eq-del-cf-ein}
    \delta\bar S_{ab}\ddot=4\hn_{(a}\tau_{b)}-\hn^c\hnabla_c\tau_{ab}-\hg_{ab}\hn^c\tau_c+(2\omega+3)\hn_a\hn_b\chi\hvarphi.
\end{equation}
At the same time, by its definition, the variation of $\bar S_{ab}$ is \cite{Wald:1999wa}
\begin{equation}
    \label{eq-del-sch}
    \delta \bar S_{ab}\ddot=-\hn_{(a}\hnabla_{b)}\tau-\hn^c\hnabla_c\tau_{ab}+\hn_{(a}\hnabla^c\tau_{b)c}+\hn_{(a}\tau_{b)}-\frac{1}{3}\hg_{ab}(\hn^c\tau_c-\hn^c\hnabla_c\tau).
\end{equation}
Comparing these two expressions, one finds out that 
\begin{subequations}
   \begin{gather}
       \hnabla^b\tau_{ab}-\hnabla_a\tau-3\tau_a-(2\omega+3)\chi\hvarphi\hn_a\ddot=0,\label{eq-vn-1}\\ 
       \hn^a\hnabla_a\tau+2\hn^a\tau_a\ddot=0.
   \end{gather} 
\end{subequations}
Because of Eq.~\eqref{eq-vn-1}, the presymplectic potential 3-form \eqref{eq-th-cf} is finite at $\scri$.
Then, the Noether charge 2-form is 
\begin{equation}
    \label{eq-noec-2-cf}
    \tilde Q_{ab}(\xi)=-\frac{1}{16\pi\tG}\bar\epsilon_{abcd}\hnabla^c(\Omega^{-2}\xi^d),
\end{equation}
which takes the same form as the integrand of Eq.~(7) in Ref.~\cite{Geroch:1981ut}, as expected.
Again, this 2-form seems to diverge at $\scri$ even worse than Eq.~\eqref{eq-th-cf}, but it is also finite there, as proved in Appendix~\ref{app-fin}.
Finally, after some tedious algebraic manipulations, the symplectic current 3-form is given by 
\begin{equation}
    \label{eq-ow-cf}
    \tilde\omega_{abc}=-\frac{1}{32\pi\tG}\bar\epsilon_{abc}(\tau'^{de}\delta N_{de}-\tau^{de}\delta N'_{de})+\frac{2\omega+3}{16\pi\tG}\bar\epsilon_{abc}(\chi'\delta \bar N-\chi\delta\bar N'),
\end{equation}
where $\tau'_{ab}$ is defined for $\delta'\hg_{ab}$, and $\chi'=\delta'\hvarphi$.
Since $\tau_{ab}=2\delta\Sigma_{ab}$, the form of this symplectic current suggests that $\sigma_{ab}$ and $N_{ab}$ are canonically conjugate to each other, so are $\hvarphi$ and $\bar N$.
From the above equation, one may choose a presymplectic potential current, given by
\begin{equation}
    \label{eq-def-Th}
    \tilde\Theta_{abc}(\delta\tg,\delta\tvarphi)=-\frac{1}{32\pi \tG}\bar\epsilon_{abc}\tau^{de} N_{de}+\frac{2\omega+3}{16\pi\tG}\bar\epsilon_{abc}\chi \bar N,
\end{equation}
so that the pullback of $\tilde\omega_{abc}$ to $\scri$ is $\delta\tilde\Theta_{abc}(\delta'\tg,\delta'\tvarphi)-\delta'\tilde{\Theta}_{abc}(\delta\tg,\delta\tvarphi)$.
There is also an ambiguity in $\tilde\Theta_{abc}$, but one may claim this is the unique one following the argument of Ref.~\cite{Wald:1999wa}.
$\tilde\Theta_{abc}$ enables the computation of the flux as discussed below.

\subsection{Fluxes}
\label{sec-flux}

Once $\tilde\Theta_{abc}$ is determined, a flux through a patch $\mathscr B$, a subset of $\scri$, can be obtained as follows
\begin{equation}
    \label{eq-def-flux}
    \begin{split}
    F_{\xi,\mathscr B}=&\int_{\mathscr B}\tilde\Theta_{abc}(\lie_\xi\tg,\lie_\xi\tvarphi)\\
        =&-\frac{1}{16\pi\tG}\int_{\mathscr B}\bar\epsilon_{abc}\Big\{  N_{de}[(\lie_\xi\mathscr D_p-\mathscr D_p\lie_\xi)\ell_q+2\ell_{(p}\mathscr D_{q)}\hK]\gamma^{dp}\gamma^{eq}\\
        &-(2\omega+3)(\lie_\xi\hvarphi+\hK\hvarphi)\bar N\Big\},
    \end{split}
\end{equation}
where $\tau_{ab}$ and $\chi$ in Eq.~\eqref{eq-def-Th} are given by $2\bar X_{ab}$ [as in Eqs.~\eqref{eq-def-bms} and \eqref{eq-x-vd}] and $\lie_\xi\hvarphi+\hK\hvarphi$ [refer to Eq.~\eqref{eq-def-vp}], respectively.  
This should be compared with Eq.~(4.14) in Ref.~\cite{Ashtekar:1981bq}, which does not contain the term with $\hvarphi$.
Suppose $\mathscr B$ is bounded by two cross sections $\mathscr C_1$ and $\mathscr C_2$ with the later in the future of the former,  then one has 
\begin{equation}
    \label{eq-c-f}
    F_{\xi,\mathscr B}=-(Q_\xi[\mathscr C_2]-Q_\xi[\mathscr C_1]).
\end{equation}
This expresses the conservation of the charge, and is also called the flux-balance law.
The overall negative sign above indicates that as the GW escapes from $\scri$, the charge of the spacetime decreases.

If $\mathscr B$ is replaced by $\scri$ in Eq.~\eqref{eq-def-flux} and the resultant integral is finite, $\mathcal H_\xi\equiv F_{\xi,\scri}$ is the Hamiltonian generator on the radiative phase space on $\scri$ associated with $\xi^a$ \cite{Bonga:2019bim}.
And using transformations \eqref{eq-def-vp}, \eqref{eq-var-sig-2}, and \eqref{eq-var-sig-lt}, one gets the Hamiltonian generators for the supertranslation $\alpha\hn^a$ and the Lorentz generator parameterized by $Y^a$, 
\begin{subequations}
    \begin{gather}
        \mathcal H_\alpha=\frac{1}{16\pi\tG}\int_{\mathscr I}\bar\epsilon_{abc}\left[N_{de}\left( \mathscr D_p\mathscr D_q\alpha+\frac{\alpha}{2}N_{pq} \right)\gamma^{dp}\gamma^{eq}+\alpha(2\omega+3)\bar N^2\right],\label{eq-h-al}\\ 
        \begin{split}
            \mathcal H_Y=\frac{1}{16\pi\tG}\int_{\mathscr I}\bar\epsilon_{abc} &\bigg\{ N_{de}\left[ \frac{u}{2}\mathscr D_p\mathscr D_q(\mathscr D\cdot Y)+\frac{1}{2}\sigma_{pq}\script D\cdot Y-\lie_Y\sigma_{pq}+\frac{u}{4}N_{pq}\script D\cdot Y \right] \\
            &\gamma^{dp}\gamma^{eq}+(2\omega+3)\bar N\left[ \frac{1}{2}(u\bar N+\hvarphi)\script D\cdot Y +\lie_Y\hvarphi \right]\bigg\},
        \end{split}
    \end{gather}
\end{subequations}
respectively.
In GR, the term linear in $N_{ab}$ in Eq.~\eqref{eq-h-al} gives the soft charge and the one quadratic in $N_{ab}$ the hard charge \cite{Ashtekar:2018lor,Alessio:2019cae}.
So by analogue, the terms linear in $N_{ab}$ and $\bar N$ determine the soft fluxes, and those quadratic in $N_{ab}$ and $\bar N$ the hard fluxes \footnote{In the terminology of Refs.~\cite{Ashtekar:2018lor,Alessio:2019cae}, $\mathcal H_\alpha$ and $\mathcal H_Y$ are both called charges, as they are given by the integrals over a 3-dimensional hypersurface, just like the electric charge: $\int_\Sigma J^dn_d\epsilon_{abc}$ where $J^a$ is the 4-current, $\Sigma$ is a spacelike hypersurface with a unit normal $n_a$ and the volume element $\epsilon_{abc}$. 
But we will call them fluxes because of Eq.~\eqref{eq-c-f}. }.
Using the results presented in Sec.~\ref{sec-bd}, one can explicitly compute the Hamiltonian generators,
\begin{subequations}
    \begin{gather}
        \mathcal H_\alpha=\frac{\varphi_0}{16\pi G_0}\int\alpha\left[ \mathscr D_A\mathscr D_BN^{AB}+\frac{1}{2}N_A^BN^A_B+(2\omega+3)\left( \frac{N}{\varphi_0} \right)^2 \right]\ud u\ud^2\boldsymbol{\Omega},\\ 
        \begin{split}
        \mathcal H_Y=\mathcal H_{\alpha'}+\frac{\varphi_0}{32\pi G_0}\int  Y^A&\left[ \frac{1}{2}(\hat c_B^C\mathscr D_AN^B_C-N_B^C\mathscr D_A\hat c^B_C)+\mathscr D^B(N^C_B\hat c_{AC}-\hat c^C_B N_{AC})\right.\\ 
            &\left.+\frac{2\omega+3}{\varphi_0^2}(N\mathscr D_A\varphi_1-\varphi_1\mathscr D_AN)\right]\ud u\ud^2\boldsymbol{\Omega},
        \end{split}
    \end{gather}
\end{subequations}
where $\alpha'=\frac{u}{2}\mathscr D_A Y^A$, $\ud^2\boldsymbol{\Omega}=\sin\theta\ud\theta\ud\phi$, and the integration by parts has been applied.
These results are consistent with those in Ref.~\cite{Tahura:2020vsa}.

\subsection{``Conserved charges''}
\label{sec-cc}

Now, it is ready to calculate the ``conserved charges''.
According to the decomposition \eqref{eq-xi-dec}, any BMS generator $\xi^a$ contains a component tangent to a cross section $\mathscr C$, and a component transverse to $\mathscr C$, once a foliation of $\scri$ is prescribed.
The ``conserved charges'' for different components will be calculated in different ways.
So one would like to rewrite $\xi^a=\xi_1^a+\xi_a^2$ with \cite{Flanagan:2015pxa}
\begin{subequations}
    \begin{gather}
        \xi^a_1\ddot=\frac{u-u_0}{2}\psi(\pd_u)^a+Y^A(\pd_A)^a,\\ 
        \xi^a_2\ddot=\left( \alpha+\frac{u_0}{2}\psi \right)(\pd_u)^a,
    \end{gather}
\end{subequations}
where $u_0$ labels some reference cross section $\mathscr C_0$, so that $\xi^a_1$ is tangent to $\script C_0$ at $u=u_0$.
These expressions imply that $\xi_1^a$ is an infinitesimal Lorentz transformation, and $\xi_2^a$ is a supertranslation generator.
The charges on $\mathscr C_0$ will be determined.

For the Lorentz generator $\xi_1^a$, the ``conserved charge'' on $\mathscr C_0$ is given by \cite{Wald:1999wa},
\begin{equation}
    \label{eq-noc-c}
    Q_{\xi_1}[\mathscr C_0]=\oint_{\mathscr C_0}\tilde Q_{ab}(\xi_1),
\end{equation}
with the requirement that $\tnabla_a\xi^a=\order{\Omega^2}$ \cite{Flanagan:2015pxa}.
This requirement is satisfied by $\xi^a$ obtained in Ref.~\cite{Hou:2020tnd}.
In order to calculate this, we employ the asymptotic solutions presented in Sec.~\ref{sec-bd} to get,
\begin{equation}
    \label{eq-noec-y}
    Q_{\xi_1}[\mathscr C_0]=\frac{1}{16\pi \tG}\oint_{\mathscr C_0}Y^A\left[ 2N_A+\frac{1}{16}\mathscr D_A(\hat c_{BC}\hat c^{BC})+\frac{2\omega+3}{4}\frac{\varphi_1\mathscr D_A\varphi_1}{\varphi_0^2} \right]\ud^2\boldsymbol{\Omega}.
\end{equation}
For the supertranslation generator $\xi^a_2$, the ``conserved charge'' satisfies \cite{Wald:1999wa}
\begin{equation}
    \label{eq-noc-ge} 
    \delta Q_{\xi_2}[\mathscr C_0]=\oint_{\mathscr C_0}[\tilde Q_{ab}(\alpha\hn)-\alpha\hn^c\tilde\theta_{cab}+\alpha\hn^c\tilde\Theta_{cab}].
\end{equation}
Unfortunately, it is very difficult to calculate this expression directly.
Instead, one can take the advantage of Eq.~\eqref{eq-c-f}.
Now, let $\xi^a=\alpha\hn^a$, so the  flux for this generator is 
\begin{equation}
    \label{eq-ev-flux}
    F_{\alpha\hn,\mathscr B}=-\frac{1}{16\pi\tG}\int_{\mathscr B}\bar\epsilon_{abc}\left[\gamma^{df}\gamma^{eh} N_{fh}\left(\lie_{\alpha\hn}\mathscr D_f-\mathscr D_f\lie_{\alpha\hn}\right)\ell_h+(2\omega+3)\alpha \bar N^2\right],
\end{equation}
by Eq.~\eqref{eq-def-flux}.
From this, one applies the Stokes' theorem to obtain the ``conserved charge'' for $\xi^a=\alpha \hn^a$ \cite{Ashtekar:1981bq},
\begin{equation}
    \label{eq-def-cgs}
    Q_\alpha[\mathscr C]=\frac{1}{8\pi\tG}\oint_{\mathscr C}P^d\ell_d\hn^c\bar\epsilon_{cab},
\end{equation}
with 
\begin{equation}
    \label{eq-def-p}
    P^a=\frac{\alpha}{4}K^{ab}\ell_b+ N_{cd}\gamma^{bd}\gamma^{c[e}\hn^{a]}(\alpha\mathscr D_e\ell_b+\ell_e\mathscr D_b\alpha)-\frac{2\omega+3}{6}\alpha\hn^a\hvarphi \bar N.
\end{equation}
By setting $\bar N=0$ in Eqs.~\eqref{eq-ev-flux} and \eqref{eq-def-p}, one recovers GR's results \cite{Geroch1977,Ashtekar:1981bq}.
Now, using the results in Sec.~\ref{sec-bd}, one obtains the ``conserved charge'' conjugate to $\xi_2^a$,
\begin{equation}
    \label{eq-cgs-v}
    Q_{\xi_2}[\mathscr C_0]=\frac{1}{8\pi \tG}\oint_{\mathscr C_0}\left( 2\alpha m-u_0Y^A\mathscr D_A m\right)\ud^2\boldsymbol{\Omega}.
\end{equation}
To obtain this expression, one has replaced $\alpha$ by $\alpha+u_0\psi/2$ in Eq.~\eqref{eq-def-cgs}.

The total ``conserved charge'' is the sum of Eqs.~\eqref{eq-noec-y} and \eqref{eq-cgs-v},
\begin{equation}
    \label{eq-tot-c}
        Q_\xi[\mathscr C]=\frac{\varphi_0}{8\pi G_0}\oint_{\mathscr C}\left[ 2\alpha m-u\lie_Ym +Y^AN_A+\frac{1}{32}\lie_Y(\hat c_{B}^A\hat c^{B}_A)+\frac{2\omega+3}{8}\frac{\varphi_1\lie_Y\varphi_1}{\varphi_0^2}\right]\ud^2\boldsymbol{\Omega},
\end{equation}
which is evaluated at some arbitrary $\mathscr C$, and is consistent with Eq.~(3.5) in Ref.~\cite{Flanagan:2015pxa}.
Of course, in the above computation, we implicitly assume that the charges of the Minkowski spacetime all vanish, as one can always add any constant to $Q_\xi$ without breaking Eq.~\eqref{eq-def-dh}. 
This imposes a nontrivial condition \cite{Wald:1999wa}, 
\begin{equation}
    \label{eq-wz-36}
    \int_{\pd\Sigma}\left\{\eta^c\tilde\theta_{cab}(\lie_\xi\tg,\lie_\xi\tvarphi)-\xi^c\tilde\theta_{cab}(\lie_\eta\tg,\lie_\eta\tvarphi)+\tilde{\mathcal L}\tepsilon_{abcd}\eta^c\xi^d-\tilde Q_{ab}[\lie_\eta\xi]\right\}=0,
\end{equation}
where $\eta^a$ is also a BMS generator, and $\tilde{\mathcal L}$ is the Lagrange density in Eq.~\eqref{eq-act-e}. In addition, $\tg_{ab}=\eta_{ab}$ and $\tvarphi=\tvarphi_0$, which are implicitly included in this expression.
One can show that this condition is satisfied as presented in Appendix~\ref{app-v36}.

Now, let us work out the ``conserved charges'' for some specific BMS generators. 
First, consider a generic supertranslation generator $\alpha\hn^a$ with $\lie_{\hn}\alpha=0$. 
The ``conserved charge'' is called the supermomentum, given by 
\begin{equation}
    \label{eq-smc}
    \mathcal P_\alpha[\mathscr C]=\frac{\varphi_0}{4\pi G_0}\oint_{\mathscr C}\alpha m\ud^2\boldsymbol{\Omega}.
\end{equation}
Among these supermomenta, four of them are special, obtained by replacing $\alpha$ by $l=0,1$ spherical harmonics.
They constitute the Bondi 4-momentum $P^a$. 
In particular, the zeroth component $P^0$ is the Bondi mass,
\begin{equation}
    \label{eq-bondi-m}
        M=\frac{\varphi_0}{4\pi G_0}\oint_{\mathscr C} m\ud^2\boldsymbol{\Omega},
\end{equation} 
which justifies the name of $m$.
In some literature, ``supermomenta'' do not include $P^a$ \cite{McCarthy1975bms,Barnich:2015uva,Flanagan:2015pxa}.
Second, switch off $\alpha$ and write $Y^A$ in the following way \cite{Flanagan:2015pxa},
\begin{equation}
    \label{eq-y-dec}
    Y^A=\mathscr D^A\mu+\epsilon^{AB}\mathscr D_B\upsilon,
\end{equation}
where $\epsilon^{AB}$ is the totally antisymmetric tensor on the unit 2-sphere, and $\mu$ and $\upsilon$ are linear combinations of $l=1$ spherical harmonics, satisfying $(\mathscr D^2+2)\mu=(\mathscr D^2+2)\upsilon=0$.
$\mu$ is the electric part and $\upsilon$ the magnetic part of $Y^A$.
The electric part generates the Lorentz boost, whose charge is 
\begin{equation}
    \label{eq-bt}
    \mathcal K_\mu[\mathscr C]=-\frac{\varphi_0}{8\pi G_0}\oint_{\mathscr C}\mu\left( \mathscr D^AN_A+2um-\frac{\hat c_A^{B}\hat c^A_B}{16}-\frac{2\omega+3}{8}\frac{\varphi_1^2}{\varphi_0^2} \right)\ud^2\boldsymbol{\Omega},
\end{equation}
and the magnetic part generates the rotation with the following charge,
\begin{equation}
    \label{eq-rt}
    \mathcal J_\upsilon[\mathscr C]=-\frac{\varphi_0}{8\pi G_0}\oint_{\mathscr C}\upsilon\epsilon^{AB}\mathscr D_AN_B\ud^2\boldsymbol{\Omega},
\end{equation}
which explains why $N_A$ is called the angular momentum aspect.
$\mathcal K_\mu$ and $\mathcal J_\upsilon$ are called the CM and the spin charges, respectively.
Since there are three linearly independent $l=1$ spherical harmonics, there are both three linearly independent boost and rotation charges. 
In total, there are six, consistent with the fact that the Lorentz algebra is six dimensional.
One should also note that the scalar field only contributes  to the boost charge $\mathcal K_\mu$.
A remark regarding the forms of the spin and CM charges is in order. 
There are different conventions in defining what is called the Bondi angular momentum aspect \cite{Strominger2014bms,Compere:2018ylh}. 
So the spin and CM charges, and the relevant fluxes, take different forms. 
These differences are summarized in Ref.~\cite{Compere:2019gft} in GR.

\section{Memories}
\label{sec-mem}

As discussed in Ref.~\cite{Hou:2020tnd}, GWs in both the tensor and the scalar sectors induce the displacement memory effects. 
There, the focus was on the relation between the memory effects and the asymptotic symmetries that induce the vacuum transitions.
Here, we will reanalyze the memory effects, concentrating on the constraints on memories imposed by the flux-balance laws. 
We will not only consider the displacement memory, but also the spin and the CM memory effects \cite{Pasterski:2015tva,Nichols:2018qac}.

We will also consider the memory effects between vacuum states in the tensor and the scalar sectors. 
Following Ref.~\cite{Hou:2020tnd}, a vacuum state in the scalar sector is simply given by $N=\dot\varphi_1=0$.
However, a vacuum state in the tensor sector is not only determined by $N_{AB}=-\pd_u\hat c_{AB}=0$, but also by the vanishing of the Newman-Penrose variables \cite{Newman:1961qr} $\Psi_4,\,\Psi_3$ and $\Psi_2-\bar\Psi_2$ at leading orders in $1/r$ \footnote{$\bar\Psi_2$ is the complex conjugate to $\Psi_2$.}.
This definition agrees with the one in GR, and also with the requirement $N_{ab}={}^*K^{ab}=0$ \cite{Ashtekar:1981bq}.
Now, write $\hat c_{AB}$ in the following way \cite{Flanagan:2015pxa},
\begin{equation}
    \label{eq-c-dec}
    \hat c_{AB}=\left(\mathscr D_A\mathscr D_B-\frac{1}{2}\gamma_{AB}\mathscr D^2\right)\Phi+\epsilon_{C(A}\mathscr D_{B)}\mathscr D^C\Upsilon,
\end{equation}
where $\Phi$ is the electric part, and $\Upsilon$ the magnetic part.
In vacuum, $\Upsilon=0$.

\subsection{Displacement memory effects}
\label{sec-dpm}

One starts with the displacement memory effect in the tensor sector. 
Rewrite the flux-balance law associated with the supertranslation $\alpha\hn^a$ in the generalized Bondi-Sachs coordinates,
\begin{equation}
    \label{eq-f-al}
    \begin{split}
    F_{\alpha\hn,\mathscr B}=&\frac{\varphi_0}{16\pi G_0}\int_{\mathscr B}\alpha\left[ \mathscr D^A\mathscr D^BN_{AB}+\frac{N_{AB}N^{AB}}{2}+(2\omega+3)\left( \frac{N}{\varphi_0} \right)^2 \right]\ud u\ud^2\boldsymbol{\Omega}\\
       =&-\Delta\mathcal P_\alpha, 
    \end{split}
\end{equation}
where $\Delta\mathcal P_\alpha=\mathcal P_{\alpha}[\mathscr C_2]-\mathcal P_{\alpha}[\mathscr C_1]$ for simplicity.
One can perform the retarded time integral of the soft flux above, and then rearrange the expression to get
\begin{equation}
    \label{eq-f-al-2}
    \oint_{\mathscr C}\alpha\mathscr D^2(\mathscr D^2+2)\Delta\Phi\ud^2\boldsymbol{\Omega}=\frac{32\pi G_0}{\varphi_0}(\mathscr E_\alpha+\Delta\mathcal P_\alpha).
\end{equation}
where $\mathscr E_\alpha$ is $F_{\alpha\hn,\mathscr B}$ without the first term in the square brackets, and Eq.~\eqref{eq-c-dec} has been used.
So $\Delta\Phi$ fully captures the displacement memory in the tensor sector, and it is completely constrained by the above equation.
One often states $\mathscr E_\alpha$ causes the null memory, and $\Delta\mathcal P_\alpha$ the ordinary memory \cite{Bieri:2013ada}.

Now, consider the displacement memory effect in the scalar sector.
Following the above argument, one may want to consider the flux-balance law for the Lorentz generator $Y^A$, given by 
\begin{equation}
    \label{eq-f-y}
    \begin{split}
        F_{Y,\mathscr B}=&\Delta\mathcal P_{\alpha'}-\frac{\varphi_0}{16\pi G_0}\int_{\mathscr B}\bigg[Y^AJ_A
        +\upsilon\epsilon_{AB}N^{CA}\hat c_C^B\bigg]\ud u\ud^2\boldsymbol{\Omega}\\
        &+\frac{\varphi_0}{64\pi G_0}\oint_{\mathscr C}\mu\Delta\left( \frac{2\omega+3}{\varphi_0^2}\varphi_1^2+\frac{\hat c_A^B\hat c^A_B}{2} \right)\ud^2\boldsymbol{\Omega}\\
        =&-\Delta\mathcal K_\mu-\Delta\mathcal J_\upsilon,
    \end{split}
\end{equation}
where $\Delta\mathcal P_{\alpha'}$ is the integral of Eq.~\eqref{eq-f-al} with $\alpha$ replaced by $\alpha'=u\mathscr D^2\mu/2=-u\mu$,  $\Delta\mathcal K_\mu$ and $\Delta\mathcal J_\upsilon$ are defined similarly to $\Delta\mathcal P_\alpha$, and 
\begin{equation}
    \label{eq-def-j}
    J_A=\frac{1}{2} N^B_C\mathscr D_A\hat c^C_B-\frac{2\omega+3}{\varphi_0^2}N\mathscr D_A\varphi_1.
\end{equation}
It is worthwhile to point it out that $\Delta\mathcal P_{\alpha'}$ is not a flux, as $\alpha'$ depends on $u$.
In fact, one could set the magnetic part $\upsilon=0$, and rearrange the expression to obtain,
\begin{equation}
    \label{eq-f-mu}
    \oint\mu\Delta\left[ \frac{2\omega+3}{\varphi_0^2}\varphi_1^2+\frac{\hat c_A^B\hat c^A_B}{2} \right]\ud^2\boldsymbol{\Omega}=-\frac{16\pi G_0}{\varphi_0}(\Delta\mathcal K_\mu+\Delta\mathcal P_{\alpha'}+\mathscr J_\mu),
\end{equation}
where $\mathscr J_\mu$ is given by 
\begin{equation}
    \label{eq-def-jmu}
    \mathscr J_\mu=\frac{\varphi_0}{16\pi G_0}\int_{\mathscr B}\mu \mathscr D^AJ_A\ud u\ud^2\boldsymbol{\Omega}.
\end{equation}
It may seems that this is a constraint equation on $\Delta\varphi_1^2$, but not, due to Eq.~\eqref{eq-bt}. 
In fact, the left hand side is canceled by the terms in $\Delta\mathcal K_\mu$.
Nevertheless, Eq.~\eqref{eq-f-y} is useful for CM memory.

In fact, the equation of motion gives a  constraint on $\Delta\varphi_1^2$, which is 
\begin{equation}
    \label{eq-con-sm}
    \Delta\varphi_1^2=\frac{16\varphi_0^2}{2\omega+3}\left\{\frac{1}{32}\Delta(\hat c_A^B\hat c_B^A)+ \mathscr D^{-2}\mathscr D^A\Delta N_A 
    -\int_{u_i}^{u_f}\ud u\left[ m+\frac{1}{2}\mathscr D^{-2}\mathscr D^AJ_A \right]\right\},
\end{equation}
where $\mathscr D^{-2}$ is the inverse operator of $\mathscr D^2$ and is explicitly given in Ref.~\cite{Hou:2020tnd}.
These results suggest that $\Delta\varphi_1$ is a persistent variable \cite{Flanagan:2018yzh} as stated in Ref.~\cite{Tahura:2020vsa}.

\subsection{Spin memory effect}
\label{sec-sm}

Spin memory effect exists only in the tensor sector, as it depends on the leading order term in $g_{uA}$ \cite{Pasterski:2015tva,Hou:2020tnd}. 
In order to determine the constraint on spin memory effect from the flux-balance law, one needs to consider the extended BMS algebra, which includes all $Y^A$ satisfying the conformal relation $\lie_Y\gamma_{AB}=\gamma_{AB}\mathscr D\cdot Y$. 
These $Y^A$ may not be globally smooth on the unit 2-sphere \cite{Blumenhagen:2009zz,Barnich:2010eb,Flanagan:2015pxa}.
However, in Sec.~\ref{sec-cc-f}, we assumed $Y^A$ are smooth vector fields, so the fluxes and charges calculated there cannot be directly used here. 
Fortunately, there is a simple remedy.
One still uses the fluxes and charges defined above,  examines the flux-balance law, finds the discrepancy and fixes it.
It turns out that without modifying the definition of the charge, one may want to add to the flux associated with $Y^A$ the following correction \cite{Flanagan:2015pxa},
\begin{equation}
    \label{eq-f-c}
    \begin{split}
    \mathcal F_{Y,\mathscr B}=&\frac{\varphi_0}{32\pi G_0}\int_{\mathscr B}Y^A\mathscr D^B(\mathscr D_A\mathscr D_C\hat c^C_B-\mathscr D_B\mathscr D_C\hat c^C_A)\ud u\ud^2\boldsymbol{\Omega}\\
        =&\frac{\varphi_0}{64\pi G_0}\int_{\mathscr B}\epsilon_{AB}Y^A\mathscr D^B\mathscr D^2(\mathscr D^2+2)\Upsilon\ud u\ud^2\boldsymbol{\Omega}.
    \end{split}
\end{equation}
This correction vanishes when $Y^A$ is smooth on the unit 2-sphere.
Now, add this term to the right hand side of the first line in Eq.~\eqref{eq-f-y}, and set $\mu=0$.
Then, one obtains the constraint on the spin memory, measured by $\Delta\mathcal R=\int\ud u\Upsilon$ \cite{Flanagan:2015pxa},
\begin{equation}
    \label{eq-f-sc}
    \oint_{\mathscr C}\upsilon\mathscr D^2\mathscr D^2(\mathscr D^2+2)\Delta\mathcal R\ud^2\boldsymbol{\Omega}=-\frac{32\pi G_0}{\varphi_0}(\Delta\mathcal J_\upsilon+\mathscr Q_\upsilon+\mathscr{\bar{J}}_\upsilon),
\end{equation}
where one has
\begin{subequations}
    \begin{gather}
        \mathscr Q_\upsilon=-\frac{\varphi_0}{16\pi G_0}\int_{\mathscr B}\upsilon\epsilon_{AB}N^{AC}\hat c^B_C\ud u\ud^2\boldsymbol{\Omega},\\
        \mathscr{\bar{J}}_\upsilon=\frac{\varphi_0}{16\pi G_0}\int_{\mathscr B}\upsilon\epsilon^{AB}\mathscr D_AJ_B\ud u\ud^2\boldsymbol{\Omega}.\label{eq-def-ju}
    \end{gather}
\end{subequations}
Note that here, $\upsilon$ is not necessarily a linear combination of $l=1$ spherical harmonics.

\subsection{Center-of-mass memory effect}
\label{sec-cmm}

Now, consider the CM memory.
Since $\mathscr D^2(\mathscr D^2+2)$ in Eq.~\eqref{eq-f-al-2} is linear, one may define $\Phi=\Phi_n+\Phi_o$ such that 
\begin{subequations}
    \begin{gather}
    \oint_{\mathscr C}\alpha\mathscr D^2(\mathscr D^2+2)\Delta\Phi_n\ud^2\boldsymbol{\Omega}=\frac{32\pi G_0}{\varphi_0}\mathscr E_\alpha,\\
    \oint_{\mathscr C}\alpha\mathscr D^2(\mathscr D^2+2)\Delta\Phi_o\ud^2\boldsymbol{\Omega}=\frac{32\pi G_0}{\varphi_0}\Delta\mathcal P_\alpha.
    \end{gather}
\end{subequations}
Then the CM memory effect is determined by \cite{Nichols:2018qac,Tahura:2020vsa}
\begin{equation}
    \label{eq-def-cmm}
    \Delta\mathscr K=\int_{u_i}^{u_f}u\pd_u\Phi_o\ud u.
\end{equation}
It appears in $\Delta\mathcal P_{\alpha'}$, i.e.,
\begin{equation}
    \label{eq-pp-k}
    \Delta\mathcal P_{\alpha'}=-\frac{\varphi_0}{64\pi G_0}\oint_{\mathscr C}\mu\mathscr D^2\mathscr D^2(\mathscr D^2+2)\Delta\mathscr K\ud^2\boldsymbol{\Omega}.
\end{equation}
So Eq.~\eqref{eq-f-y} can be rewritten to yield,
\begin{equation}
    \label{eq-f-y-mu}
    \oint_{\mathscr C}\mu\mathscr D^2\mathscr D^2(\mathscr D^2+2)\Delta\mathscr K\ud^2\boldsymbol{\Omega}=\frac{64\pi G_0}{\varphi_0}(\mathscr J_\mu-\Delta\mathcal K'_\mu),
\end{equation}
where $\Delta\mathcal K'_\mu$ is not the change in any charge, given by
\begin{equation}
    \label{eq-def-kp}
    \Delta\mathcal K'_\mu=-\frac{\varphi_0}{8\pi G_0}\oint_{\mathscr C}\mu\Delta(\mathscr D^AN_A+2um)\ud^2\boldsymbol{\Omega}.
\end{equation}
Therefore, the CM memory is constrained by Eq.~\eqref{eq-f-y-mu}, as long as $\mu$ is not simply a linear combination of $l=1$ spherical harmonics.

\section{Conclusion}
\label{sec-con}

In this work, we analyzed the asymptotic structure and the BMS symmetries in an isolated system in BD using the covariant conformal completion method. 
The results thus obtained are independent of the coordinate system used. 
There are also four different orders of asymptotic structure as in GR. 
The zeroth-order structure $(\gamma_{ab},\hn^a)$ is universal, and the first-order structure $\{\mathscr D_a\}$ characterizes the differences among spacetimes.
The second-order structure $(N_{ab},\bar N)$ is the radiative degrees of freedom, and the third-order structure ${}^*K^{ab}$ contains the full gauge covariant information in $\{\mathscr D_a\}$ \cite{Ashtekar:2014zsa}.
The BMS symmetries also include the supertranslations and the Lorentz transformations, and their actions on the asymptotic structure are discussed.
Based on these, the ``conserved charges'' and fluxes are computed with Wald-Zoupas formulism. 
If one switches off the scalar field, one reproduces GR's results. 
The scalar field only contributes to the CM charge, but it appears in all fluxes.
Finally, the flux-balance laws are used to constrain various memory effects. 
Among them, the displacement memory effect in the scalar sector cannot be restricted by the flux-balance laws, but the equation of motion constrains it partially.
Memory effects in the tensor sector are well constrained by the flux-balance laws as in GR.

\begin{acknowledgements}
This work was supported by the National Natural Science Foundation of China under grant Nos.~11633001 and 11920101003, and the Strategic Priority Research Program of the Chinese Academy of Sciences, grant No.~XDB23000000.
This was also a project funded by China Postdoctoral Science Foundation (No.~2020M672400).
\end{acknowledgements}

\appendix

\section{Finiteness of the Noether charge}
\label{app-fin}

In this section,  one shows the finiteness of Eq.~\eqref{eq-noec-2-cf}. 
Because of Stokes' theorem, one writes
\begin{equation}
    \label{eq-noec-v}
    \int_{\mathscr C}\tilde Q_{ab}=\int_{S_0}\tilde Q_{ab}+\int_{\Sigma'}\ud_a\tilde Q_{bc}
\end{equation}
where $S_0$ is a finite topological 2-sphere in the physical spacetime, and $\Sigma'$ is a 3-dimensional hypersurface joining $S_0$ to $\mathscr C$.
If the last two integrals are both finite, then the first is also finite, and thus, so is the integrand $\tilde Q_{ab}$.

The integrand of the third integral can be contracted with $\bar\epsilon^{abcd}$ and one can thus examine \cite{Geroch:1981ut} 
\begin{equation}
    \label{eq-ig-e}
\hnabla^b\hnabla_{[a}(\Omega^{-2}\xi_{b]})=\Omega^{-2}\bar R_{ab}\xi^b+\hnabla_a\hnabla_b(\Omega^{-2}\xi^b)-\hnabla^b\hnabla_{(a}(\Omega^{-2}\xi_{b)}), 
\end{equation}
with $\bar R_{ab}$ the Ricci tensor of $\hg_{ab}$.
Using  the property \eqref{eq-def-bms}, one can reexpress the last two terms in the above equation in the following way,
\begin{equation}
    \label{eq-ig-e-2}
    \begin{split}
\hnabla_a\hnabla_b(\Omega^{-2}\xi^b)-\hnabla^b\hnabla_{(a}(\Omega^{-2}\xi_{b)})=&\Omega^{-1}(3X_a+\hnabla_aX-\hnabla^bX_{ab})\\
       &+\Omega^{-3}\xi^b(2\hnabla_a\hn_b+\hg_{ab}\hnabla_c\hn^c-3\Omega^{-1}\hg_{ab}\hn_c\hn^c). 
    \end{split}
\end{equation}
Then, by Eq.~\eqref{eq-cf-ein}, one knows that 
\begin{equation}
    \label{eq-cfein-2}
    \Omega\bar R_{ab}+2\hnabla_a\hn_b+\hg_{ab}\hnabla_c\hn^c-2\Omega^{-1}\hg_{ab}\hn_c\hn^c-\frac{1}{2}\Omega^{-1}\hg_{ab}\bar L=\Omega^{-1}\bar L_{ab},
\end{equation}
with $\bar L=\hg^{ab}\bar L_{ab}$.
Therefore, Eq.~\eqref{eq-ig-e} becomes
\begin{equation}
    \label{eq-ig-e-f}
\hnabla^b\hnabla_{[a}(\Omega^{-2}\xi_{b]})=\Omega^{-1}(3X_a+\hnabla_aX-\hnabla^bX_{ab})+\Omega^{-4}\xi^b\left( \bar L_{ab}+\frac{1}{2}\hg_{ab}\bar L \right).
\end{equation}
Now, substitute in the definition \eqref{eq-def-l} of $\bar L_{ab}$, and then 
\begin{equation}
    \label{eq-ig-e-f1}
    \begin{split}
\hnabla^b\hnabla_{[a}(\Omega^{-2}\xi_{b]})=&\Omega^{-1}\left[  3X_a+\hnabla_aX-\hnabla^bX_{ab}+\frac{2\omega+3}{2}(\hK\hvarphi^2+\hvarphi\lie_\xi\hvarphi)\hn_a\right]\\
&+\hK\hvarphi\hnabla_a\hvarphi+\hnabla_a\hvarphi\lie_\xi\hvarphi.
    \end{split}
\end{equation}
Again, Eq.~\eqref{eq-ig-e-f1} on $\scri$ seems to be also diverge, which is not true.
To understand this, one wants to use Eq.~\eqref{eq-glinkage-3}, and for that, one has to calculate
\begin{subequations}
    \begin{gather}
        \begin{split}
        \lie_\xi\bar R_{ab}=&-2\hnabla_a\hnabla_b\hK+2\hn_{(a}(X_{b)}-\hnabla_{b)}X+\hnabla^cX_{b)c})+4X_{c(a}\hnabla_{b)}\hn^c-X_{ab}\hnabla_c\hn^c\\ 
        &-X\hnabla_a\hn_b-2\lie_{\hn}X_{ab}+\Omega(2\hnabla_c\hnabla_{(a}X^c_{b)}+2\hnabla_{(a}X_{b)}-\hnabla_c\hnabla^cX_{ab}-\hnabla_a\hnabla_bX),
        \end{split}\\ 
        \begin{split}
            \lie_\xi\bar R=&-2\hnabla_a\hnabla^a\hK+4\hn^aX_a-2X_{ab}\hnabla^a\hn^b-2X\hnabla_a\hn^a-4\lie_{\hn}X-2\hK\bar R\\ 
            &+2\Omega(\hnabla_a\hnabla_bX^{ab}-\hnabla_a\hnabla^aX+2\hnabla_aX^a-X^{ab}\bar R_{ab}).
        \end{split}\label{eq-lie-xi-rs}
    \end{gather}
\end{subequations}
One also has to know $\lie_\xi(\Omega^{-2}\bar L_{ab})$, which can be checked to be $\order{\Omega}$.
Therefore, Eq.~\eqref{eq-glinkage-3} leads to
\begin{equation}
    \label{eq-glk-3-f}
    \begin{split}
    &\hn_{(a}\left[3X_{b)}+\hnabla_{b)}-\hnabla^cX_{b)c}+\frac{2\omega+3}{2}\hvarphi(\lie_\xi\hvarphi+\hK\hvarphi)\hn_{b)}\right]\\ &-\frac{1}{6}\hg_{ab}(\hnabla_c\hnabla^c\hK+4\hn^cX_c+2\lie_{\hn}X)\ddot=0.
    \end{split}
\end{equation}
It turns out that $\hnabla_a\hnabla^a\hK=0$.
Therefore, one knows that 
\begin{subequations}
    \begin{gather}
        3X_a+\hnabla_aX-\hnabla^bX_{ab}+\frac{2\omega+3}{2}\hvarphi(\lie_\xi\hvarphi+\hK\hvarphi)\hn_a\ddot=0,\\ 
        2\hn^aX_a+\lie_{\hn}X\ddot=0.
    \end{gather}
\end{subequations}
This implies that Eq.~\eqref{eq-ig-e-f1} is indeed finite, so is Eq.~\eqref{eq-noec-v}.
Therefore, Eq.~\eqref{eq-noec-2-cf} is finite on $\scri$.

\section{Verify condition~\eqref{eq-wz-36}}
\label{app-v36}

One has to check that Eq.~\eqref{eq-wz-36} should hold in BD. 
In Minkowski spacetime, any BMS generator is a sum of a supertranslation and a Killing vector field.
If either $\eta^a$ or $\xi^a$ is a Killing vector field, Eq.~\eqref{eq-wz-36} is satisfied \cite{Wald:1999wa}.
As in GR, one only has to check if
\begin{equation}
    \label{eq-wz-36-1}
    \int_{\pd\Sigma}[\eta^c\tilde\theta_{cab}(\lie_\xi\tg,\lie_\xi\tvarphi)-\xi^c\tilde\theta_{cab}(\lie_\eta\tg,\lie_\eta\tvarphi)]=0,
\end{equation}
where $\xi^a\ddot=\alpha\hn^a$ and $\eta^a\ddot=\beta\hn^a$ are two supertranslation generators with $\lie_{\hn}\alpha=\lie_{\hn}\beta=0$.
So let us calculate the first term in the square brackets above, which is
\begin{equation}
    \label{eq-wz-76}
    \eta^c\tilde\theta_{cab}=\frac{1}{16\pi\tG}\beta F(\alpha)\hn^c\bar\epsilon_{cab},
\end{equation}
with the function $F(\alpha)$ given by 
\begin{equation}
    \label{eq-wz-77}
    \begin{split}
    F(\alpha)=&-\hn_a\Omega^{-1}[\hnabla_b\lambda^{ab}-\hnabla^a\lambda-3\lambda^a-(2\omega+3)\chi\hvarphi\hn^a]+(2\omega+3)\chi\bar N\\
       \ddot=&-\hnabla_a\hnabla_b\lambda^{ab}+\hnabla^2\lambda+3\hnabla_a\lambda^a+(2\omega+3)\chi\bar N,
    \end{split}
\end{equation}
where $\lambda_{ab}=\Omega\lie_\xi\tg_{ab}=\Omega^{-1}(\lie_\xi\hg_{ab}-2\hK\hg_{ab})$, $\lambda=\hg^{ab}\lambda_{ab}$ and $\lambda_a=\lambda_{ab}\hn^b/\Omega$.
The first three terms add up to a quantity proportional to the so-called ``flux'' defined by Eq.~(19) in Ref.~\cite{Geroch:1981ut}, when the gauge condition $\tnabla_a\xi^a=0$ is imposed. 
As discussed in that work, their flux can also be calculates using their Eq.~(20), which is gauge invariant.
So in the current case, we can also rewrite the above expression,
\begin{equation}
    \label{eq-gw-20}
    F(\alpha)\ddot=-\hnabla_a\hnabla_b\lambda^{ab}+3\hnabla_a\lambda^a+\frac{3}{4}\hnabla^2\lambda+\frac{1}{24}\bar R\lambda+(2\omega+3)\chi\bar N.
\end{equation} 
Now, one should calculate $F(\alpha)$ with $\xi^a=\alpha\hn^a-\Omega\hnabla^a\alpha$ \cite{Geroch:1981ut}, then, one obtains that 
\begin{gather}
  \hK=\Omega(\alpha\vartheta-\sigma_\alpha),\\
  \lambda_{ab}=-2\bar\nabla_a\bar\nabla_b\alpha-(\alpha\vartheta-2\sigma_\alpha)\bar g_{ab}-\alpha (\bar S_{ab}-\Omega^{-2}\bar L_{ab}),\\
  \lambda_a=\bar\nabla_a(\alpha\vartheta-2\sigma_\alpha)-(\bar S_{ab}-\Omega^{-2}\bar L_{ab})\bar\nabla^b\alpha,\\
  \lambda=-2\bar\nabla^2\alpha-4(\alpha\vartheta-2\sigma_\alpha)-\alpha\left(\frac{\bar R}{3}-\Omega^{-2}\bar L  \right).
\end{gather}
With these, one finds out that 
\begin{equation}
    \label{eq-gw-20-1}
    F(\alpha)\ddot=-\frac{\alpha(2\omega+3)}{3}\left[2\bar N^2-\hvarphi\lie_{\hn}^2\hvarphi\right]-\frac{1}{4}\left(\hnabla^2-\frac{\bar R}{6}\right)\left(\hnabla^2\alpha+2\alpha\vartheta-4\sigma_\alpha+\frac{\alpha \bar R}{6}\right),
\end{equation}
where $\lie_{\hn}^2\hvarphi=\lie_{\hn}\bar N$.
One can easily verify that 
\begin{equation}
    \label{eq-lpal}
    \hnabla^2\alpha\ddot=\mathscr D^2\alpha+2\sigma_\alpha.
\end{equation}
In order to calculate $\hnabla^2\hnabla^2\alpha$, one needs 
\begin{equation}
    \label{eq-ln-lpal}
    \lie_{\hn}\hnabla^2\alpha=
  2\lie_{\hn}\sigma_\alpha+\Omega\left( \bar\nabla^2\sigma_\alpha+2\vartheta\sigma_\alpha-\frac{\bar R\sigma_\alpha}{3}+\frac{1}{6}\bar\nabla_a\alpha\bar\nabla^a\bar R-\vartheta\bar{\nabla}^2\alpha+\bar S^{ab}\bar\nabla_a \bar\nabla_b\alpha\right)+\order{\Omega^2},
\end{equation}
where $\order{\Omega^n}$ means some finite term at $\scri$ multiplied by $\Omega^n$.
So one has
\begin{equation}
    \label{eq-lplpal}
    \hnabla^2\hnabla^2\alpha=\mathscr D^2\mathscr D^2\alpha+4\bar\nabla^2\sigma_\alpha+4\vartheta\sigma_\alpha-\frac{2}{3}\bar R\sigma_\alpha+\frac{1}{3}\bar\nabla_a\alpha\bar\nabla^a\bar R-2\vartheta\bar\nabla^2\alpha+2\bar S^{ab}\bar\nabla_a\bar\nabla_b\alpha+\order{\Omega}.
\end{equation}
To proceed further, one needs $\lie_{\hn}\vartheta$ and $\lie_{\hn}\bar R$.
Then Eq.~\eqref{eq-lie-xi-rs} is useful, and one sets $\xi^a=\hn^a$ there.
So one finds out that 
\begin{gather}
    X^{\hn}_{ab}=-\frac{\vartheta}{2}\hg_{ab}-\frac{1}{2}(\bar S_{ab}-\Omega^{-2}\bar L_{ab}),\quad X^{\hn}_a=\frac{\hnabla_a\vartheta}{2},\\
    X^{\hn}=-2\vartheta-\frac{\bar R}{6}+\frac{\Omega^{-2}\bar L}{2}, \quad \hK^{\hn}=\Omega\vartheta.
\end{gather}
As we know that, $\hnabla^2\hK^{\hn}=0$, one gets
\begin{equation}
    \label{eq-lpk}
    \hnabla^2\hK^{\hn}=2\lie_{\hn}\vartheta+\Omega\left(\hnabla^2\vartheta+ 2\vartheta^2-\frac{\vartheta\bar R}{6} \right)+\order{\Omega^2},
\end{equation}
so 
\begin{equation}
    \label{eq-lien-g}
    \lie_{\hn}\vartheta=-\Omega\left(\frac{1}{2}\hnabla^2\vartheta+ \vartheta^2-\frac{\vartheta\bar R}{12} \right)+\order{\Omega^2}.
\end{equation}
This implies that 
\begin{equation}
    \label{eq-lpg}
    \hnabla^2\vartheta=-\vartheta^2+\frac{\vartheta\bar R}{12}+\frac{1}{2}\mathscr D^2\vartheta+\order{\Omega}.
\end{equation}
Equation~\eqref{eq-lie-xi-rs} gives
\begin{equation}
    \label{eq-lxi-rs-v}
    \lie_{\hn}\bar R=\Omega\left\{-\frac{\vartheta\bar R}{2}+\frac{3}{2}\bar S^{ab}\bar S_{ab}+(2\omega+3)\left[ 4\bar N^2+\hvarphi\lie_{\hn}^2\hvarphi \right]\right\}+\order{\Omega^2},
\end{equation}
and thus,
\begin{equation}
    \label{eq-lprs}
    \hnabla^2\bar R=-\vartheta\bar R+3\bar S^{ab}\bar S_{ab}+2(2\omega+3)\left[ 4\bar N^2+\hvarphi\lie_{\hn}^2\hvarphi \right]+\order{\Omega}.
\end{equation}
Finally, the ``flux'' is given by 
\begin{equation}
    \label{eq-flux-f}
    F(\alpha)=\frac{1}{2}\left(\mathscr D^2\mathscr D^2\alpha+2\mathscr D^2\alpha+2 N^{ab}\mathscr D_a\mathscr D_b\alpha+\frac{\alpha}{2} N^{ab} N_{ab} \right)+\frac{2(2\omega+3)}{3}\alpha\lie_{\hn}(\hvarphi\bar N).
\end{equation}
In Minkowski spacetime, $N_{ab}=0$ and $\bar N=0$, therefore, 
\begin{equation}
    \label{eq-flux-m}
F(\alpha)=\frac{1}{2}\left(\mathscr D^2\mathscr D^2\alpha+2\mathscr D^2\alpha  \right),
\end{equation}
which implies that Eq.~\eqref{eq-wz-36} is satisfied.
$F(\alpha)$ is exactly the same to the one in GR up to a factor \cite{Ashtekar1982lk}.

\bibliographystyle{apsrev4-1}
\bibliography{ConservedChargesST.bbl}

\begin{thebibliography}{74}%
\makeatletter
\providecommand \@ifxundefined [1]{%
 \@ifx{#1\undefined}
}%
\providecommand \@ifnum [1]{%
 \ifnum #1\expandafter \@firstoftwo
 \else \expandafter \@secondoftwo
 \fi
}%
\providecommand \@ifx [1]{%
 \ifx #1\expandafter \@firstoftwo
 \else \expandafter \@secondoftwo
 \fi
}%
\providecommand \natexlab [1]{#1}%
\providecommand \enquote  [1]{``#1''}%
\providecommand \bibnamefont  [1]{#1}%
\providecommand \bibfnamefont [1]{#1}%
\providecommand \citenamefont [1]{#1}%
\providecommand \href@noop [0]{\@secondoftwo}%
\providecommand \href [0]{\begingroup \@sanitize@url \@href}%
\providecommand \@href[1]{\@@startlink{#1}\@@href}%
\providecommand \@@href[1]{\endgroup#1\@@endlink}%
\providecommand \@sanitize@url [0]{\catcode `\\12\catcode `\$12\catcode
  `\&12\catcode `\#12\catcode `\^12\catcode `\_12\catcode `\%12\relax}%
\providecommand \@@startlink[1]{}%
\providecommand \@@endlink[0]{}%
\providecommand \url  [0]{\begingroup\@sanitize@url \@url }%
\providecommand \@url [1]{\endgroup\@href {#1}{\urlprefix }}%
\providecommand \urlprefix  [0]{URL }%
\providecommand \Eprint [0]{\href }%
\providecommand \doibase [0]{http://dx.doi.org/}%
\providecommand \selectlanguage [0]{\@gobble}%
\providecommand \bibinfo  [0]{\@secondoftwo}%
\providecommand \bibfield  [0]{\@secondoftwo}%
\providecommand \translation [1]{[#1]}%
\providecommand \BibitemOpen [0]{}%
\providecommand \bibitemStop [0]{}%
\providecommand \bibitemNoStop [0]{.\EOS\space}%
\providecommand \EOS [0]{\spacefactor3000\relax}%
\providecommand \BibitemShut  [1]{\csname bibitem#1\endcsname}%
\let\auto@bib@innerbib\@empty
\bibitem [{\citenamefont {Abbott}\ \emph
  {et~al.}(2016{\natexlab{a}})\citenamefont {Abbott} \emph
  {et~al.}}]{Abbott:2016blz}%
  \BibitemOpen
  \bibfield  {author} {\bibinfo {author} {\bibfnamefont {B.~P.}\ \bibnamefont
  {Abbott}} \emph {et~al.} (\bibinfo {collaboration} {Virgo, LIGO
  Scientific}),\ }\href {\doibase 10.1103/PhysRevLett.116.061102} {\bibfield
  {journal} {\bibinfo  {journal} {Phys. Rev. Lett.}\ }\textbf {\bibinfo
  {volume} {116}},\ \bibinfo {pages} {061102} (\bibinfo {year}
  {2016}{\natexlab{a}})},\ \Eprint {http://arxiv.org/abs/1602.03837}
  {arXiv:1602.03837 [gr-qc]} \BibitemShut {NoStop}%
\bibitem [{\citenamefont {Abbott}\ \emph
  {et~al.}(2016{\natexlab{b}})\citenamefont {Abbott} \emph
  {et~al.}}]{Abbott:2016nmj}%
  \BibitemOpen
  \bibfield  {author} {\bibinfo {author} {\bibfnamefont {B.~P.}\ \bibnamefont
  {Abbott}} \emph {et~al.} (\bibinfo {collaboration} {Virgo, LIGO
  Scientific}),\ }\href {\doibase 10.1103/PhysRevLett.116.241103} {\bibfield
  {journal} {\bibinfo  {journal} {Phys. Rev. Lett.}\ }\textbf {\bibinfo
  {volume} {116}},\ \bibinfo {pages} {241103} (\bibinfo {year}
  {2016}{\natexlab{b}})},\ \Eprint {http://arxiv.org/abs/1606.04855}
  {arXiv:1606.04855 [gr-qc]} \BibitemShut {NoStop}%
\bibitem [{\citenamefont {Abbott}\ \emph
  {et~al.}(2017{\natexlab{a}})\citenamefont {Abbott} \emph
  {et~al.}}]{Abbott:2017vtc}%
  \BibitemOpen
  \bibfield  {author} {\bibinfo {author} {\bibfnamefont {B.~P.}\ \bibnamefont
  {Abbott}} \emph {et~al.} (\bibinfo {collaboration} {Virgo, LIGO
  Scientific}),\ }\href {\doibase 10.1103/PhysRevLett.118.221101} {\bibfield
  {journal} {\bibinfo  {journal} {Phys. Rev. Lett.}\ }\textbf {\bibinfo
  {volume} {118}},\ \bibinfo {pages} {221101} (\bibinfo {year}
  {2017}{\natexlab{a}})},\ \Eprint {http://arxiv.org/abs/1706.01812}
  {arXiv:1706.01812 [gr-qc]} \BibitemShut {NoStop}%
\bibitem [{\citenamefont {Abbott}\ \emph
  {et~al.}(2017{\natexlab{b}})\citenamefont {Abbott} \emph
  {et~al.}}]{Abbott:2017oio}%
  \BibitemOpen
  \bibfield  {author} {\bibinfo {author} {\bibfnamefont {B.~P.}\ \bibnamefont
  {Abbott}} \emph {et~al.} (\bibinfo {collaboration} {Virgo, LIGO
  Scientific}),\ }\href {\doibase 10.1103/PhysRevLett.119.141101} {\bibfield
  {journal} {\bibinfo  {journal} {Phys. Rev. Lett.}\ }\textbf {\bibinfo
  {volume} {119}},\ \bibinfo {pages} {141101} (\bibinfo {year}
  {2017}{\natexlab{b}})},\ \Eprint {http://arxiv.org/abs/1709.09660}
  {arXiv:1709.09660 [gr-qc]} \BibitemShut {NoStop}%
\bibitem [{\citenamefont {Abbott}\ \emph
  {et~al.}(2017{\natexlab{c}})\citenamefont {Abbott} \emph
  {et~al.}}]{TheLIGOScientific:2017qsa}%
  \BibitemOpen
  \bibfield  {author} {\bibinfo {author} {\bibfnamefont {B.~P.}\ \bibnamefont
  {Abbott}} \emph {et~al.} (\bibinfo {collaboration} {Virgo, LIGO
  Scientific}),\ }\href {\doibase 10.1103/PhysRevLett.119.161101} {\bibfield
  {journal} {\bibinfo  {journal} {Phys. Rev. Lett.}\ }\textbf {\bibinfo
  {volume} {119}},\ \bibinfo {pages} {161101} (\bibinfo {year}
  {2017}{\natexlab{c}})},\ \Eprint {http://arxiv.org/abs/1710.05832}
  {arXiv:1710.05832 [gr-qc]} \BibitemShut {NoStop}%
\bibitem [{\citenamefont {Abbott}\ \emph
  {et~al.}(2017{\natexlab{d}})\citenamefont {Abbott} \emph
  {et~al.}}]{Abbott:2017gyy}%
  \BibitemOpen
  \bibfield  {author} {\bibinfo {author} {\bibfnamefont {B.~P.}\ \bibnamefont
  {Abbott}} \emph {et~al.} (\bibinfo {collaboration} {Virgo and LIGO Scientific
  Collaborations}),\ }\href {\doibase 10.3847/2041-8213/aa9f0c} {\bibfield
  {journal} {\bibinfo  {journal} {Astrophys. J.}\ }\textbf {\bibinfo {volume}
  {851}},\ \bibinfo {pages} {L35} (\bibinfo {year} {2017}{\natexlab{d}})},\
  \Eprint {http://arxiv.org/abs/1711.05578} {arXiv:1711.05578 [astro-ph.HE]}
  \BibitemShut {NoStop}%
\bibitem [{\citenamefont {Abbott}\ \emph {et~al.}(2019)\citenamefont {Abbott}
  \emph {et~al.}}]{LIGOScientific:2018mvr}%
  \BibitemOpen
  \bibfield  {author} {\bibinfo {author} {\bibfnamefont {B.~P.}\ \bibnamefont
  {Abbott}} \emph {et~al.} (\bibinfo {collaboration} {Virgo and LIGO Scientific
  Collaborations}),\ }\href {\doibase 10.1103/PhysRevX.9.031040} {\bibfield
  {journal} {\bibinfo  {journal} {Phys. Rev. X}\ }\textbf {\bibinfo {volume}
  {9}},\ \bibinfo {pages} {031040} (\bibinfo {year} {2019})},\ \Eprint
  {http://arxiv.org/abs/1811.12907} {arXiv:1811.12907 [astro-ph.HE]}
  \BibitemShut {NoStop}%
\bibitem [{\citenamefont {Abbott}\ \emph
  {et~al.}(2020{\natexlab{a}})\citenamefont {Abbott} \emph
  {et~al.}}]{Abbott:2020uma}%
  \BibitemOpen
  \bibfield  {author} {\bibinfo {author} {\bibfnamefont {B.~P.}\ \bibnamefont
  {Abbott}} \emph {et~al.} (\bibinfo {collaboration} {Virgo and LIGO Scientific
  Collaborations}),\ }\href {\doibase 10.3847/2041-8213/ab75f5} {\bibfield
  {journal} {\bibinfo  {journal} {Astrophys.\ J.\ Lett.}\ }\textbf {\bibinfo
  {volume} {892}},\ \bibinfo {pages} {L3} (\bibinfo {year}
  {2020}{\natexlab{a}})},\ \Eprint {http://arxiv.org/abs/2001.01761}
  {arXiv:2001.01761 [astro-ph.HE]} \BibitemShut {NoStop}%
\bibitem [{\citenamefont {Abbott}\ \emph
  {et~al.}(2020{\natexlab{b}})\citenamefont {Abbott} \emph
  {et~al.}}]{LIGOScientific:2020stg}%
  \BibitemOpen
  \bibfield  {author} {\bibinfo {author} {\bibfnamefont {R.}~\bibnamefont
  {Abbott}} \emph {et~al.} (\bibinfo {collaboration} {LIGO Scientific,
  Virgo}),\ }\href {\doibase 10.1103/PhysRevD.102.043015} {\bibfield  {journal}
  {\bibinfo  {journal} {Phys. Rev. D}\ }\textbf {\bibinfo {volume} {102}},\
  \bibinfo {pages} {043015} (\bibinfo {year} {2020}{\natexlab{b}})},\ \Eprint
  {http://arxiv.org/abs/2004.08342} {arXiv:2004.08342 [astro-ph.HE]}
  \BibitemShut {NoStop}%
\bibitem [{\citenamefont {Abbott}\ \emph
  {et~al.}(2020{\natexlab{c}})\citenamefont {Abbott} \emph
  {et~al.}}]{Abbott:2020khf}%
  \BibitemOpen
  \bibfield  {author} {\bibinfo {author} {\bibfnamefont {R.}~\bibnamefont
  {Abbott}} \emph {et~al.} (\bibinfo {collaboration} {LIGO Scientific,
  Virgo}),\ }\href {\doibase 10.3847/2041-8213/ab960f} {\bibfield  {journal}
  {\bibinfo  {journal} {Astrophys. J. Lett.}\ }\textbf {\bibinfo {volume}
  {896}},\ \bibinfo {pages} {L44} (\bibinfo {year} {2020}{\natexlab{c}})},\
  \Eprint {http://arxiv.org/abs/2006.12611} {arXiv:2006.12611 [astro-ph.HE]}
  \BibitemShut {NoStop}%
\bibitem [{\citenamefont {Abbott}\ \emph
  {et~al.}(2020{\natexlab{d}})\citenamefont {Abbott} \emph
  {et~al.}}]{Abbott:2020tfl}%
  \BibitemOpen
  \bibfield  {author} {\bibinfo {author} {\bibfnamefont {R.}~\bibnamefont
  {Abbott}} \emph {et~al.} (\bibinfo {collaboration} {LIGO Scientific,
  Virgo}),\ }\href {\doibase 10.1103/PhysRevLett.125.101102} {\bibfield
  {journal} {\bibinfo  {journal} {Phys. Rev. Lett.}\ }\textbf {\bibinfo
  {volume} {125}},\ \bibinfo {pages} {101102} (\bibinfo {year}
  {2020}{\natexlab{d}})},\ \Eprint {http://arxiv.org/abs/2009.01075}
  {arXiv:2009.01075 [gr-qc]} \BibitemShut {NoStop}%
\bibitem [{\citenamefont {Abbott}\ \emph
  {et~al.}(2020{\natexlab{e}})\citenamefont {Abbott} \emph
  {et~al.}}]{Abbott:2020mjq}%
  \BibitemOpen
  \bibfield  {author} {\bibinfo {author} {\bibfnamefont {R.}~\bibnamefont
  {Abbott}} \emph {et~al.} (\bibinfo {collaboration} {LIGO Scientific,
  Virgo}),\ }\href {\doibase 10.3847/2041-8213/aba493} {\bibfield  {journal}
  {\bibinfo  {journal} {Astrophys. J. Lett.}\ }\textbf {\bibinfo {volume}
  {900}},\ \bibinfo {pages} {L13} (\bibinfo {year} {2020}{\natexlab{e}})},\
  \Eprint {http://arxiv.org/abs/2009.01190} {arXiv:2009.01190 [astro-ph.HE]}
  \BibitemShut {NoStop}%
\bibitem [{\citenamefont {Abbott}\ \emph
  {et~al.}(2020{\natexlab{f}})\citenamefont {Abbott} \emph
  {et~al.}}]{Abbott:2020niy}%
  \BibitemOpen
  \bibfield  {author} {\bibinfo {author} {\bibfnamefont {R.}~\bibnamefont
  {Abbott}} \emph {et~al.} (\bibinfo {collaboration} {LIGO Scientific,
  Virgo}),\ }\href@noop {} {\  (\bibinfo {year} {2020}{\natexlab{f}})},\
  \Eprint {http://arxiv.org/abs/2010.14527} {arXiv:2010.14527 [gr-qc]}
  \BibitemShut {NoStop}%
\bibitem [{\citenamefont {Einstein}(1916)}]{Einstein:1916cc}%
  \BibitemOpen
  \bibfield  {author} {\bibinfo {author} {\bibfnamefont {A.}~\bibnamefont
  {Einstein}},\ }\href@noop {} {\bibfield  {journal} {\bibinfo  {journal}
  {Sitzungsber. Preuss. Akad. Wiss. Berlin (Math. Phys.)}\ }\textbf {\bibinfo
  {volume} {1916}},\ \bibinfo {pages} {688} (\bibinfo {year}
  {1916})}\BibitemShut {NoStop}%
\bibitem [{\citenamefont {Einstein}(1918)}]{Einstein:1918btx}%
  \BibitemOpen
  \bibfield  {author} {\bibinfo {author} {\bibfnamefont {A.}~\bibnamefont
  {Einstein}},\ }\href@noop {} {\bibfield  {journal} {\bibinfo  {journal}
  {Sitzungsber. Preuss. Akad. Wiss. Berlin (Math. Phys.)}\ }\textbf {\bibinfo
  {volume} {1918}},\ \bibinfo {pages} {154} (\bibinfo {year}
  {1918})}\BibitemShut {NoStop}%
\bibitem [{\citenamefont {Will}(2014)}]{Will:2014kxa}%
  \BibitemOpen
  \bibfield  {author} {\bibinfo {author} {\bibfnamefont {C.~M.}\ \bibnamefont
  {Will}},\ }\href {\doibase 10.12942/lrr-2014-4} {\bibfield  {journal}
  {\bibinfo  {journal} {Living Rev. Rel.}\ }\textbf {\bibinfo {volume} {17}},\
  \bibinfo {pages} {4} (\bibinfo {year} {2014})},\ \Eprint
  {http://arxiv.org/abs/1403.7377} {arXiv:1403.7377 [gr-qc]} \BibitemShut
  {NoStop}%
\bibitem [{\citenamefont {{Gong}}\ and\ \citenamefont
  {{Hou}}(2018)}]{Gong:2018ybk}%
  \BibitemOpen
  \bibfield  {author} {\bibinfo {author} {\bibfnamefont {Y.}~\bibnamefont
  {{Gong}}}\ and\ \bibinfo {author} {\bibfnamefont {S.}~\bibnamefont {{Hou}}},\
  }\href {\doibase 10.3390/universe4080085} {\bibfield  {journal} {\bibinfo
  {journal} {Universe}\ }\textbf {\bibinfo {volume} {4}},\ \bibinfo {pages}
  {85} (\bibinfo {year} {2018})},\ \Eprint {http://arxiv.org/abs/1806.04027}
  {arXiv:1806.04027 [gr-qc]} \BibitemShut {NoStop}%
\bibitem [{\citenamefont {Zel'dovich}\ and\ \citenamefont
  {Polnarev}(1974)}]{Zeldovich:1974gvh}%
  \BibitemOpen
  \bibfield  {author} {\bibinfo {author} {\bibfnamefont {Y.~B.}\ \bibnamefont
  {Zel'dovich}}\ and\ \bibinfo {author} {\bibfnamefont {A.~G.}\ \bibnamefont
  {Polnarev}},\ }\href@noop {} {\bibfield  {journal} {\bibinfo  {journal} {Sov.
  Astron.}\ }\textbf {\bibinfo {volume} {18}},\ \bibinfo {pages} {17} (\bibinfo
  {year} {1974})}\BibitemShut {NoStop}%
\bibitem [{\citenamefont {Braginsky}\ and\ \citenamefont
  {Grishchuk}(1985)}]{Braginsky:1986ia}%
  \BibitemOpen
  \bibfield  {author} {\bibinfo {author} {\bibfnamefont {V.}~\bibnamefont
  {Braginsky}}\ and\ \bibinfo {author} {\bibfnamefont {L.}~\bibnamefont
  {Grishchuk}},\ }\href@noop {} {\bibfield  {journal} {\bibinfo  {journal}
  {Sov. Phys. JETP}\ }\textbf {\bibinfo {volume} {62}},\ \bibinfo {pages} {427}
  (\bibinfo {year} {1985})}\BibitemShut {NoStop}%
\bibitem [{\citenamefont {Christodoulou}(1991)}]{Christodoulou1991}%
  \BibitemOpen
  \bibfield  {author} {\bibinfo {author} {\bibfnamefont {D.}~\bibnamefont
  {Christodoulou}},\ }\href {\doibase 10.1103/PhysRevLett.67.1486} {\bibfield
  {journal} {\bibinfo  {journal} {Phys. Rev. Lett.}\ }\textbf {\bibinfo
  {volume} {67}},\ \bibinfo {pages} {1486} (\bibinfo {year}
  {1991})}\BibitemShut {NoStop}%
\bibitem [{\citenamefont {Bondi}\ \emph {et~al.}(1962)\citenamefont {Bondi},
  \citenamefont {van~der Burg},\ and\ \citenamefont {Metzner}}]{Bondi:1962px}%
  \BibitemOpen
  \bibfield  {author} {\bibinfo {author} {\bibfnamefont {H.}~\bibnamefont
  {Bondi}}, \bibinfo {author} {\bibfnamefont {M.~G.~J.}\ \bibnamefont {van~der
  Burg}}, \ and\ \bibinfo {author} {\bibfnamefont {A.~W.~K.}\ \bibnamefont
  {Metzner}},\ }\href {\doibase 10.1098/rspa.1962.0161} {\bibfield  {journal}
  {\bibinfo  {journal} {Proc. Roy. Soc. Lond. A}\ }\textbf {\bibinfo {volume}
  {269}},\ \bibinfo {pages} {21} (\bibinfo {year} {1962})}\BibitemShut
  {NoStop}%
\bibitem [{\citenamefont {Sachs}(1962{\natexlab{a}})}]{Sachs:1962wk}%
  \BibitemOpen
  \bibfield  {author} {\bibinfo {author} {\bibfnamefont {R.~K.}\ \bibnamefont
  {Sachs}},\ }\href {\doibase 10.1098/rspa.1962.0206} {\bibfield  {journal}
  {\bibinfo  {journal} {Proc. Roy. Soc. Lond. A}\ }\textbf {\bibinfo {volume}
  {270}},\ \bibinfo {pages} {103} (\bibinfo {year}
  {1962}{\natexlab{a}})}\BibitemShut {NoStop}%
\bibitem [{\citenamefont {Sachs}(1962{\natexlab{b}})}]{Sachs:1962zza}%
  \BibitemOpen
  \bibfield  {author} {\bibinfo {author} {\bibfnamefont {R.}~\bibnamefont
  {Sachs}},\ }\href {\doibase 10.1103/PhysRev.128.2851} {\bibfield  {journal}
  {\bibinfo  {journal} {Phys. Rev.}\ }\textbf {\bibinfo {volume} {128}},\
  \bibinfo {pages} {2851} (\bibinfo {year} {1962}{\natexlab{b}})}\BibitemShut
  {NoStop}%
\bibitem [{\citenamefont {Strominger}\ and\ \citenamefont
  {Zhiboedov}(2016)}]{Strominger:2014pwa}%
  \BibitemOpen
  \bibfield  {author} {\bibinfo {author} {\bibfnamefont {A.}~\bibnamefont
  {Strominger}}\ and\ \bibinfo {author} {\bibfnamefont {A.}~\bibnamefont
  {Zhiboedov}},\ }\href {\doibase 10.1007/JHEP01(2016)086} {\bibfield
  {journal} {\bibinfo  {journal} {JHEP}\ }\textbf {\bibinfo {volume} {01}},\
  \bibinfo {pages} {086} (\bibinfo {year} {2016})},\ \Eprint
  {http://arxiv.org/abs/1411.5745} {arXiv:1411.5745 [hep-th]} \BibitemShut
  {NoStop}%
\bibitem [{\citenamefont {Pasterski}\ \emph {et~al.}(2016)\citenamefont
  {Pasterski}, \citenamefont {Strominger},\ and\ \citenamefont
  {Zhiboedov}}]{Pasterski:2015tva}%
  \BibitemOpen
  \bibfield  {author} {\bibinfo {author} {\bibfnamefont {S.}~\bibnamefont
  {Pasterski}}, \bibinfo {author} {\bibfnamefont {A.}~\bibnamefont
  {Strominger}}, \ and\ \bibinfo {author} {\bibfnamefont {A.}~\bibnamefont
  {Zhiboedov}},\ }\href {\doibase 10.1007/JHEP12(2016)053} {\bibfield
  {journal} {\bibinfo  {journal} {JHEP}\ }\textbf {\bibinfo {volume} {12}},\
  \bibinfo {pages} {053} (\bibinfo {year} {2016})},\ \Eprint
  {http://arxiv.org/abs/1502.06120} {arXiv:1502.06120 [hep-th]} \BibitemShut
  {NoStop}%
\bibitem [{\citenamefont {Nichols}(2018)}]{Nichols:2018qac}%
  \BibitemOpen
  \bibfield  {author} {\bibinfo {author} {\bibfnamefont {D.~A.}\ \bibnamefont
  {Nichols}},\ }\href {\doibase 10.1103/PhysRevD.98.064032} {\bibfield
  {journal} {\bibinfo  {journal} {Phys. Rev. D}\ }\textbf {\bibinfo {volume}
  {98}},\ \bibinfo {pages} {064032} (\bibinfo {year} {2018})},\ \Eprint
  {http://arxiv.org/abs/1807.08767} {arXiv:1807.08767 [gr-qc]} \BibitemShut
  {NoStop}%
\bibitem [{\citenamefont {Lang}(2014)}]{Lang:2013fna}%
  \BibitemOpen
  \bibfield  {author} {\bibinfo {author} {\bibfnamefont {R.~N.}\ \bibnamefont
  {Lang}},\ }\href {\doibase 10.1103/PhysRevD.89.084014} {\bibfield  {journal}
  {\bibinfo  {journal} {Phys. Rev. D}\ }\textbf {\bibinfo {volume} {89}},\
  \bibinfo {pages} {084014} (\bibinfo {year} {2014})},\ \Eprint
  {http://arxiv.org/abs/1310.3320} {arXiv:1310.3320 [gr-qc]} \BibitemShut
  {NoStop}%
\bibitem [{\citenamefont {Lang}(2015)}]{Lang:2014osa}%
  \BibitemOpen
  \bibfield  {author} {\bibinfo {author} {\bibfnamefont {R.~N.}\ \bibnamefont
  {Lang}},\ }\href {\doibase 10.1103/PhysRevD.91.084027} {\bibfield  {journal}
  {\bibinfo  {journal} {Phys. Rev. D}\ }\textbf {\bibinfo {volume} {91}},\
  \bibinfo {pages} {084027} (\bibinfo {year} {2015})},\ \Eprint
  {http://arxiv.org/abs/1411.3073} {arXiv:1411.3073 [gr-qc]} \BibitemShut
  {NoStop}%
\bibitem [{\citenamefont {Du}\ and\ \citenamefont
  {Nishizawa}(2016)}]{Du:2016hww}%
  \BibitemOpen
  \bibfield  {author} {\bibinfo {author} {\bibfnamefont {S.~M.}\ \bibnamefont
  {Du}}\ and\ \bibinfo {author} {\bibfnamefont {A.}~\bibnamefont {Nishizawa}},\
  }\href {\doibase 10.1103/PhysRevD.94.104063} {\bibfield  {journal} {\bibinfo
  {journal} {Phys. Rev. D}\ }\textbf {\bibinfo {volume} {94}},\ \bibinfo
  {pages} {104063} (\bibinfo {year} {2016})},\ \Eprint
  {http://arxiv.org/abs/1609.09825} {arXiv:1609.09825 [gr-qc]} \BibitemShut
  {NoStop}%
\bibitem [{\citenamefont {Kilicarslan}\ and\ \citenamefont
  {Tekin}(2019)}]{Kilicarslan:2018bia}%
  \BibitemOpen
  \bibfield  {author} {\bibinfo {author} {\bibfnamefont {E.}~\bibnamefont
  {Kilicarslan}}\ and\ \bibinfo {author} {\bibfnamefont {B.}~\bibnamefont
  {Tekin}},\ }\href {\doibase 10.1140/epjc/s10052-019-6636-4} {\bibfield
  {journal} {\bibinfo  {journal} {Eur. Phys. J. C}\ }\textbf {\bibinfo {volume}
  {79}},\ \bibinfo {pages} {114} (\bibinfo {year} {2019})},\ \Eprint
  {http://arxiv.org/abs/1805.02240} {arXiv:1805.02240 [gr-qc]} \BibitemShut
  {NoStop}%
\bibitem [{\citenamefont {Kilicarslan}(2018)}]{Kilicarslan:2018yxd}%
  \BibitemOpen
  \bibfield  {author} {\bibinfo {author} {\bibfnamefont {E.}~\bibnamefont
  {Kilicarslan}},\ }\href {\doibase 10.1103/PhysRevD.98.064048} {\bibfield
  {journal} {\bibinfo  {journal} {Phys. Rev. D}\ }\textbf {\bibinfo {volume}
  {98}},\ \bibinfo {pages} {064048} (\bibinfo {year} {2018})},\ \Eprint
  {http://arxiv.org/abs/1808.00266} {arXiv:1808.00266 [gr-qc]} \BibitemShut
  {NoStop}%
\bibitem [{\citenamefont {Kilicarslan}(2019)}]{Kilicarslan:2018unm}%
  \BibitemOpen
  \bibfield  {author} {\bibinfo {author} {\bibfnamefont {E.}~\bibnamefont
  {Kilicarslan}},\ }\href {\doibase 10.3906/fiz-1811-2} {\bibfield  {journal}
  {\bibinfo  {journal} {Turk. J. Phys.}\ }\textbf {\bibinfo {volume} {43}},\
  \bibinfo {pages} {126} (\bibinfo {year} {2019})},\ \Eprint
  {http://arxiv.org/abs/1811.00843} {arXiv:1811.00843 [gr-qc]} \BibitemShut
  {NoStop}%
\bibitem [{\citenamefont {Hou}(2021)}]{Hou:2020xme}%
  \BibitemOpen
  \bibfield  {author} {\bibinfo {author} {\bibfnamefont {S.}~\bibnamefont
  {Hou}},\ }\href {\doibase https://doi.org/10.1002/asna.202113887} {\bibfield
  {journal} {\bibinfo  {journal} {Astronomische Nachrichten}\ }\textbf
  {\bibinfo {volume} {2021}},\ \bibinfo {pages} {1} (\bibinfo {year} {2021})},\
  \Eprint {http://arxiv.org/abs/2011.02087} {arXiv:2011.02087 [gr-qc]}
  \BibitemShut {NoStop}%
\bibitem [{\citenamefont {Hou}\ and\ \citenamefont {Zhu}(2021)}]{Hou:2020tnd}%
  \BibitemOpen
  \bibfield  {author} {\bibinfo {author} {\bibfnamefont {S.}~\bibnamefont
  {Hou}}\ and\ \bibinfo {author} {\bibfnamefont {Z.-H.}\ \bibnamefont {Zhu}},\
  }\href {\doibase 10.1007/JHEP01(2021)083} {\bibfield  {journal} {\bibinfo
  {journal} {JHEP}\ }\textbf {\bibinfo {volume} {01}},\ \bibinfo {pages} {083}
  (\bibinfo {year} {2021})},\ \Eprint {http://arxiv.org/abs/2005.01310}
  {arXiv:2005.01310 [gr-qc]} \BibitemShut {NoStop}%
\bibitem [{\citenamefont {Brans}\ and\ \citenamefont
  {Dicke}(1961)}]{Brans:1961sx}%
  \BibitemOpen
  \bibfield  {author} {\bibinfo {author} {\bibfnamefont {C.}~\bibnamefont
  {Brans}}\ and\ \bibinfo {author} {\bibfnamefont {R.}~\bibnamefont {Dicke}},\
  }\href {\doibase 10.1103/PhysRev.124.925} {\bibfield  {journal} {\bibinfo
  {journal} {Phys. Rev.}\ }\textbf {\bibinfo {volume} {124}},\ \bibinfo {pages}
  {925} (\bibinfo {year} {1961})}\BibitemShut {NoStop}%
\bibitem [{\citenamefont {Tahura}\ \emph {et~al.}(2020)\citenamefont {Tahura},
  \citenamefont {Nichols}, \citenamefont {Saffer}, \citenamefont {Stein},\ and\
  \citenamefont {Yagi}}]{Tahura:2020vsa}%
  \BibitemOpen
  \bibfield  {author} {\bibinfo {author} {\bibfnamefont {S.}~\bibnamefont
  {Tahura}}, \bibinfo {author} {\bibfnamefont {D.~A.}\ \bibnamefont {Nichols}},
  \bibinfo {author} {\bibfnamefont {A.}~\bibnamefont {Saffer}}, \bibinfo
  {author} {\bibfnamefont {L.~C.}\ \bibnamefont {Stein}}, \ and\ \bibinfo
  {author} {\bibfnamefont {K.}~\bibnamefont {Yagi}},\ }\href@noop {} {\
  (\bibinfo {year} {2020})},\ \Eprint {http://arxiv.org/abs/2007.13799}
  {arXiv:2007.13799 [gr-qc]} \BibitemShut {NoStop}%
\bibitem [{\citenamefont {Penrose}(1963)}]{Penrose:1962ij}%
  \BibitemOpen
  \bibfield  {author} {\bibinfo {author} {\bibfnamefont {R.}~\bibnamefont
  {Penrose}},\ }\href {\doibase 10.1103/PhysRevLett.10.66} {\bibfield
  {journal} {\bibinfo  {journal} {Phys.\ Rev.\ Lett.}\ }\textbf {\bibinfo
  {volume} {10}},\ \bibinfo {pages} {66} (\bibinfo {year} {1963})}\BibitemShut
  {NoStop}%
\bibitem [{\citenamefont {Penrose}(1965)}]{Penrose:1965am}%
  \BibitemOpen
  \bibfield  {author} {\bibinfo {author} {\bibfnamefont {R.}~\bibnamefont
  {Penrose}},\ }\href {\doibase 10.1098/rspa.1965.0058} {\bibfield  {journal}
  {\bibinfo  {journal} {Proc. Roy. Soc. Lond. A}\ }\textbf {\bibinfo {volume}
  {284}},\ \bibinfo {pages} {159} (\bibinfo {year} {1965})}\BibitemShut
  {NoStop}%
\bibitem [{\citenamefont {Wald}\ and\ \citenamefont
  {Zoupas}(2000)}]{Wald:1999wa}%
  \BibitemOpen
  \bibfield  {author} {\bibinfo {author} {\bibfnamefont {R.~M.}\ \bibnamefont
  {Wald}}\ and\ \bibinfo {author} {\bibfnamefont {A.}~\bibnamefont {Zoupas}},\
  }\href {\doibase 10.1103/PhysRevD.61.084027} {\bibfield  {journal} {\bibinfo
  {journal} {Phys. Rev. D}\ }\textbf {\bibinfo {volume} {61}},\ \bibinfo
  {pages} {084027} (\bibinfo {year} {2000})},\ \Eprint
  {http://arxiv.org/abs/gr-qc/9911095} {arXiv:gr-qc/9911095 [gr-qc]}
  \BibitemShut {NoStop}%
\bibitem [{\citenamefont {Chandrasekaran}\ \emph {et~al.}(2018)\citenamefont
  {Chandrasekaran}, \citenamefont {Flanagan},\ and\ \citenamefont
  {Prabhu}}]{Chandrasekaran:2018aop}%
  \BibitemOpen
  \bibfield  {author} {\bibinfo {author} {\bibfnamefont {V.}~\bibnamefont
  {Chandrasekaran}}, \bibinfo {author} {\bibfnamefont {E.~E.}\ \bibnamefont
  {Flanagan}}, \ and\ \bibinfo {author} {\bibfnamefont {K.}~\bibnamefont
  {Prabhu}},\ }\href {\doibase 10.1007/JHEP11(2018)125} {\bibfield  {journal}
  {\bibinfo  {journal} {JHEP}\ }\textbf {\bibinfo {volume} {11}},\ \bibinfo
  {pages} {125} (\bibinfo {year} {2018})},\ \Eprint
  {http://arxiv.org/abs/1807.11499} {arXiv:1807.11499 [hep-th]} \BibitemShut
  {NoStop}%
\bibitem [{\citenamefont {Bonga}\ \emph {et~al.}(2020)\citenamefont {Bonga},
  \citenamefont {Grant},\ and\ \citenamefont {Prabhu}}]{Bonga:2019bim}%
  \BibitemOpen
  \bibfield  {author} {\bibinfo {author} {\bibfnamefont {B.}~\bibnamefont
  {Bonga}}, \bibinfo {author} {\bibfnamefont {A.~M.}\ \bibnamefont {Grant}}, \
  and\ \bibinfo {author} {\bibfnamefont {K.}~\bibnamefont {Prabhu}},\ }\href
  {\doibase 10.1103/PhysRevD.101.044013} {\bibfield  {journal} {\bibinfo
  {journal} {Phys. Rev. D}\ }\textbf {\bibinfo {volume} {101}},\ \bibinfo
  {pages} {044013} (\bibinfo {year} {2020})},\ \Eprint
  {http://arxiv.org/abs/1911.04514} {arXiv:1911.04514 [gr-qc]} \BibitemShut
  {NoStop}%
\bibitem [{\citenamefont {Bhattacharya}\ and\ \citenamefont
  {Majhi}(2017)}]{Bhattacharya:2017pqc}%
  \BibitemOpen
  \bibfield  {author} {\bibinfo {author} {\bibfnamefont {K.}~\bibnamefont
  {Bhattacharya}}\ and\ \bibinfo {author} {\bibfnamefont {B.~R.}\ \bibnamefont
  {Majhi}},\ }\href {\doibase 10.1103/PhysRevD.95.064026} {\bibfield  {journal}
  {\bibinfo  {journal} {Phys. Rev. D}\ }\textbf {\bibinfo {volume} {95}},\
  \bibinfo {pages} {064026} (\bibinfo {year} {2017})},\ \Eprint
  {http://arxiv.org/abs/1702.07166} {arXiv:1702.07166 [gr-qc]} \BibitemShut
  {NoStop}%
\bibitem [{\citenamefont {Bhattacharya}\ \emph {et~al.}(2018)\citenamefont
  {Bhattacharya}, \citenamefont {Das},\ and\ \citenamefont
  {Majhi}}]{Bhattacharya:2018xlq}%
  \BibitemOpen
  \bibfield  {author} {\bibinfo {author} {\bibfnamefont {K.}~\bibnamefont
  {Bhattacharya}}, \bibinfo {author} {\bibfnamefont {A.}~\bibnamefont {Das}}, \
  and\ \bibinfo {author} {\bibfnamefont {B.~R.}\ \bibnamefont {Majhi}},\ }\href
  {\doibase 10.1103/PhysRevD.97.124013} {\bibfield  {journal} {\bibinfo
  {journal} {Phys. Rev. D}\ }\textbf {\bibinfo {volume} {97}},\ \bibinfo
  {pages} {124013} (\bibinfo {year} {2018})},\ \Eprint
  {http://arxiv.org/abs/1803.03771} {arXiv:1803.03771 [gr-qc]} \BibitemShut
  {NoStop}%
\bibitem [{\citenamefont {Duval}\ \emph
  {et~al.}(2014{\natexlab{a}})\citenamefont {Duval}, \citenamefont {Gibbons},\
  and\ \citenamefont {Horvathy}}]{Duval:2014uva}%
  \BibitemOpen
  \bibfield  {author} {\bibinfo {author} {\bibfnamefont {C.}~\bibnamefont
  {Duval}}, \bibinfo {author} {\bibfnamefont {G.}~\bibnamefont {Gibbons}}, \
  and\ \bibinfo {author} {\bibfnamefont {P.}~\bibnamefont {Horvathy}},\ }\href
  {\doibase 10.1088/0264-9381/31/9/092001} {\bibfield  {journal} {\bibinfo
  {journal} {Class. Quant. Grav.}\ }\textbf {\bibinfo {volume} {31}},\ \bibinfo
  {pages} {092001} (\bibinfo {year} {2014}{\natexlab{a}})},\ \Eprint
  {http://arxiv.org/abs/1402.5894} {arXiv:1402.5894 [gr-qc]} \BibitemShut
  {NoStop}%
\bibitem [{\citenamefont {Duval}\ \emph
  {et~al.}(2014{\natexlab{b}})\citenamefont {Duval}, \citenamefont {Gibbons},\
  and\ \citenamefont {Horvathy}}]{Duval:2014lpa}%
  \BibitemOpen
  \bibfield  {author} {\bibinfo {author} {\bibfnamefont {C.}~\bibnamefont
  {Duval}}, \bibinfo {author} {\bibfnamefont {G.}~\bibnamefont {Gibbons}}, \
  and\ \bibinfo {author} {\bibfnamefont {P.}~\bibnamefont {Horvathy}},\ }\href
  {\doibase 10.1088/1751-8113/47/33/335204} {\bibfield  {journal} {\bibinfo
  {journal} {J. Phys. A}\ }\textbf {\bibinfo {volume} {47}},\ \bibinfo {pages}
  {335204} (\bibinfo {year} {2014}{\natexlab{b}})},\ \Eprint
  {http://arxiv.org/abs/1403.4213} {arXiv:1403.4213 [hep-th]} \BibitemShut
  {NoStop}%
\bibitem [{\citenamefont {Godazgar}\ \emph
  {et~al.}(2020{\natexlab{a}})\citenamefont {Godazgar}, \citenamefont
  {Godazgar},\ and\ \citenamefont {Perry}}]{Godazgar:2020gqd}%
  \BibitemOpen
  \bibfield  {author} {\bibinfo {author} {\bibfnamefont {H.}~\bibnamefont
  {Godazgar}}, \bibinfo {author} {\bibfnamefont {M.}~\bibnamefont {Godazgar}},
  \ and\ \bibinfo {author} {\bibfnamefont {M.~J.}\ \bibnamefont {Perry}},\
  }\href {\doibase 10.1103/PhysRevLett.125.101301} {\bibfield  {journal}
  {\bibinfo  {journal} {Phys. Rev. Lett.}\ }\textbf {\bibinfo {volume} {125}},\
  \bibinfo {pages} {101301} (\bibinfo {year} {2020}{\natexlab{a}})},\ \Eprint
  {http://arxiv.org/abs/2007.01257} {arXiv:2007.01257 [hep-th]} \BibitemShut
  {NoStop}%
\bibitem [{\citenamefont {Godazgar}\ \emph
  {et~al.}(2020{\natexlab{b}})\citenamefont {Godazgar}, \citenamefont
  {Godazgar},\ and\ \citenamefont {Perry}}]{Godazgar:2020kqd}%
  \BibitemOpen
  \bibfield  {author} {\bibinfo {author} {\bibfnamefont {H.}~\bibnamefont
  {Godazgar}}, \bibinfo {author} {\bibfnamefont {M.}~\bibnamefont {Godazgar}},
  \ and\ \bibinfo {author} {\bibfnamefont {M.~J.}\ \bibnamefont {Perry}},\
  }\href {\doibase 10.1007/JHEP09(2020)084} {\bibfield  {journal} {\bibinfo
  {journal} {JHEP}\ }\textbf {\bibinfo {volume} {20}},\ \bibinfo {pages} {084}
  (\bibinfo {year} {2020}{\natexlab{b}})},\ \Eprint
  {http://arxiv.org/abs/2007.07144} {arXiv:2007.07144 [hep-th]} \BibitemShut
  {NoStop}%
\bibitem [{\citenamefont {Geroch}(1977)}]{Geroch1977}%
  \BibitemOpen
  \bibfield  {author} {\bibinfo {author} {\bibfnamefont {R.}~\bibnamefont
  {Geroch}},\ }\enquote {\bibinfo {title} {Asymptotic structure of
  space-time},}\ in\ \href {\doibase 10.1007/978-1-4684-2343-3_1} {\emph
  {\bibinfo {booktitle} {Asymptotic Structure of Space-Time}}},\ \bibinfo
  {editor} {edited by\ \bibinfo {editor} {\bibfnamefont {F.~P.}\ \bibnamefont
  {Esposito}}\ and\ \bibinfo {editor} {\bibfnamefont {L.}~\bibnamefont
  {Witten}}}\ (\bibinfo  {publisher} {Springer US},\ \bibinfo {address}
  {Boston, MA},\ \bibinfo {year} {1977})\ pp.\ \bibinfo {pages}
  {1--105}\BibitemShut {NoStop}%
\bibitem [{\citenamefont {Ashtekar}(1981)}]{Ashtekar:1981hw}%
  \BibitemOpen
  \bibfield  {author} {\bibinfo {author} {\bibfnamefont {A.}~\bibnamefont
  {Ashtekar}},\ }\href {\doibase 10.1063/1.525169} {\bibfield  {journal}
  {\bibinfo  {journal} {J. Math. Phys.}\ }\textbf {\bibinfo {volume} {22}},\
  \bibinfo {pages} {2885} (\bibinfo {year} {1981})}\BibitemShut {NoStop}%
\bibitem [{\citenamefont {Ashtekar}\ and\ \citenamefont
  {Streubel}(1981)}]{Ashtekar:1981bq}%
  \BibitemOpen
  \bibfield  {author} {\bibinfo {author} {\bibfnamefont {A.}~\bibnamefont
  {Ashtekar}}\ and\ \bibinfo {author} {\bibfnamefont {M.}~\bibnamefont
  {Streubel}},\ }\href {\doibase 10.1098/rspa.1981.0109} {\bibfield  {journal}
  {\bibinfo  {journal} {Proc. Roy. Soc. Lond. A}\ }\textbf {\bibinfo {volume}
  {376}},\ \bibinfo {pages} {585} (\bibinfo {year} {1981})}\BibitemShut
  {NoStop}%
\bibitem [{\citenamefont {Ashtekar}(2014)}]{Ashtekar:2014zsa}%
  \BibitemOpen
  \bibfield  {author} {\bibinfo {author} {\bibfnamefont {A.}~\bibnamefont
  {Ashtekar}},\ }\href@noop {} {\  (\bibinfo {year} {2014})},\ \Eprint
  {http://arxiv.org/abs/1409.1800} {arXiv:1409.1800 [gr-qc]} \BibitemShut
  {NoStop}%
\bibitem [{\citenamefont {Wald}(1984)}]{Wald:1984rg}%
  \BibitemOpen
  \bibfield  {author} {\bibinfo {author} {\bibfnamefont {R.~M.}\ \bibnamefont
  {Wald}},\ }\href {\doibase 10.7208/chicago/9780226870373.001.0001} {\emph
  {\bibinfo {title} {{General Relativity}}}}\ (\bibinfo  {publisher}
  {University of Chicago Press},\ \bibinfo {address} {Chicago, IL},\ \bibinfo
  {year} {1984})\BibitemShut {NoStop}%
\bibitem [{\citenamefont {Poisson}\ and\ \citenamefont
  {Will}(2014)}]{Poisson2014}%
  \BibitemOpen
  \bibfield  {author} {\bibinfo {author} {\bibfnamefont {E.}~\bibnamefont
  {Poisson}}\ and\ \bibinfo {author} {\bibfnamefont {C.~M.}\ \bibnamefont
  {Will}},\ }\href {\doibase 10.1017/CBO9781139507486} {\emph {\bibinfo {title}
  {Gravity: Newtonian, Post-Newtonian, Relativistic}}}\ (\bibinfo  {publisher}
  {Cambridge University Press},\ \bibinfo {year} {2014})\BibitemShut {NoStop}%
\bibitem [{\citenamefont {Barnich}\ and\ \citenamefont
  {Troessaert}(2010)}]{Barnich:2010eb}%
  \BibitemOpen
  \bibfield  {author} {\bibinfo {author} {\bibfnamefont {G.}~\bibnamefont
  {Barnich}}\ and\ \bibinfo {author} {\bibfnamefont {C.}~\bibnamefont
  {Troessaert}},\ }\href {\doibase 10.1007/JHEP05(2010)062} {\bibfield
  {journal} {\bibinfo  {journal} {JHEP}\ }\textbf {\bibinfo {volume} {05}},\
  \bibinfo {pages} {062} (\bibinfo {year} {2010})},\ \Eprint
  {http://arxiv.org/abs/1001.1541} {arXiv:1001.1541 [hep-th]} \BibitemShut
  {NoStop}%
\bibitem [{\citenamefont {Hawking}\ and\ \citenamefont
  {Ellis}(2011)}]{Hawking:1973uf}%
  \BibitemOpen
  \bibfield  {author} {\bibinfo {author} {\bibfnamefont {S.~W.}\ \bibnamefont
  {Hawking}}\ and\ \bibinfo {author} {\bibfnamefont {G.~F.~R.}\ \bibnamefont
  {Ellis}},\ }\href {\doibase 10.1017/CBO9780511524646} {\emph {\bibinfo
  {title} {{The Large Scale Structure of Space-Time}}}},\ Cambridge Monographs
  on Mathematical Physics\ (\bibinfo  {publisher} {Cambridge University
  Press},\ \bibinfo {year} {2011})\BibitemShut {NoStop}%
\bibitem [{\citenamefont {Geroch}\ and\ \citenamefont
  {Winicour}(1981)}]{Geroch:1981ut}%
  \BibitemOpen
  \bibfield  {author} {\bibinfo {author} {\bibfnamefont {R.~P.}\ \bibnamefont
  {Geroch}}\ and\ \bibinfo {author} {\bibfnamefont {J.}~\bibnamefont
  {Winicour}},\ }\href {\doibase 10.1063/1.524987} {\bibfield  {journal}
  {\bibinfo  {journal} {J. Math. Phys.}\ }\textbf {\bibinfo {volume} {22}},\
  \bibinfo {pages} {803} (\bibinfo {year} {1981})}\BibitemShut {NoStop}%
\bibitem [{Note1()}]{Note1}%
  \BibitemOpen
  \bibinfo {note} {Note that $\protect \bar {K}=\psi /2$ defined in \cite
  {Hou:2020tnd}.}\BibitemShut {Stop}%
\bibitem [{\citenamefont {Alessio}\ and\ \citenamefont
  {Arzano}(2019)}]{Alessio:2019cae}%
  \BibitemOpen
  \bibfield  {author} {\bibinfo {author} {\bibfnamefont {F.}~\bibnamefont
  {Alessio}}\ and\ \bibinfo {author} {\bibfnamefont {M.}~\bibnamefont
  {Arzano}},\ }\href {\doibase 10.1103/PhysRevD.100.044028} {\bibfield
  {journal} {\bibinfo  {journal} {Phys. Rev. D}\ }\textbf {\bibinfo {volume}
  {100}},\ \bibinfo {pages} {044028} (\bibinfo {year} {2019})},\ \Eprint
  {http://arxiv.org/abs/1906.05036} {arXiv:1906.05036 [gr-qc]} \BibitemShut
  {NoStop}%
\bibitem [{\citenamefont {Wald}(1990)}]{Wald:1990closed}%
  \BibitemOpen
  \bibfield  {author} {\bibinfo {author} {\bibfnamefont {R.~M.}\ \bibnamefont
  {Wald}},\ }\href {\doibase 10.1063/1.528839} {\bibfield  {journal} {\bibinfo
  {journal} {J. Math. Phys.}\ }\textbf {\bibinfo {volume} {31}},\ \bibinfo
  {pages} {2378} (\bibinfo {year} {1990})}\BibitemShut {NoStop}%
\bibitem [{\citenamefont {Iyer}\ and\ \citenamefont
  {Wald}(1994)}]{Iyer:1994ys}%
  \BibitemOpen
  \bibfield  {author} {\bibinfo {author} {\bibfnamefont {V.}~\bibnamefont
  {Iyer}}\ and\ \bibinfo {author} {\bibfnamefont {R.~M.}\ \bibnamefont
  {Wald}},\ }\href {\doibase 10.1103/PhysRevD.50.846} {\bibfield  {journal}
  {\bibinfo  {journal} {Phys. Rev. D}\ }\textbf {\bibinfo {volume} {50}},\
  \bibinfo {pages} {846} (\bibinfo {year} {1994})},\ \Eprint
  {http://arxiv.org/abs/gr-qc/9403028} {arXiv:gr-qc/9403028 [gr-qc]}
  \BibitemShut {NoStop}%
\bibitem [{\citenamefont {Ashtekar}\ \emph {et~al.}(2018)\citenamefont
  {Ashtekar}, \citenamefont {Campiglia},\ and\ \citenamefont
  {Laddha}}]{Ashtekar:2018lor}%
  \BibitemOpen
  \bibfield  {author} {\bibinfo {author} {\bibfnamefont {A.}~\bibnamefont
  {Ashtekar}}, \bibinfo {author} {\bibfnamefont {M.}~\bibnamefont {Campiglia}},
  \ and\ \bibinfo {author} {\bibfnamefont {A.}~\bibnamefont {Laddha}},\ }\href
  {\doibase 10.1007/s10714-018-2464-3} {\bibfield  {journal} {\bibinfo
  {journal} {Gen. Rel. Grav.}\ }\textbf {\bibinfo {volume} {50}},\ \bibinfo
  {pages} {140} (\bibinfo {year} {2018})},\ \Eprint
  {http://arxiv.org/abs/1808.07093} {arXiv:1808.07093 [gr-qc]} \BibitemShut
  {NoStop}%
\bibitem [{Note2()}]{Note2}%
  \BibitemOpen
  \bibinfo {note} {In the terminology of Refs.~\cite
  {Ashtekar:2018lor,Alessio:2019cae}, $\protect \mathcal H_\alpha $ and
  $\protect \mathcal H_Y$ are both called charges, as they are given by the
  integrals over a 3-dimensional hypersurface, just like the electric charge:
  $\DOTSI \intop \ilimits@ _\Sigma J^dn_d\epsilon _{abc}$ where $J^a$ is the
  4-current, $\Sigma $ is a spacelike hypersurface with a unit normal $n_a$ and
  the volume element $\epsilon _{abc}$. But we will call them fluxes because of
  Eq.~\protect \textup {\hbox {\mathsurround \z@ \protect \normalfont
  (\ignorespaces \ref {eq-c-f}\unskip \@@italiccorr )}}.}\BibitemShut {Stop}%
\bibitem [{\citenamefont {Flanagan}\ and\ \citenamefont
  {Nichols}(2017)}]{Flanagan:2015pxa}%
  \BibitemOpen
  \bibfield  {author} {\bibinfo {author} {\bibfnamefont {E.~E.}\ \bibnamefont
  {Flanagan}}\ and\ \bibinfo {author} {\bibfnamefont {D.~A.}\ \bibnamefont
  {Nichols}},\ }\href {\doibase 10.1103/PhysRevD.95.044002} {\bibfield
  {journal} {\bibinfo  {journal} {Phys. Rev. D}\ }\textbf {\bibinfo {volume}
  {95}},\ \bibinfo {pages} {044002} (\bibinfo {year} {2017})},\ \Eprint
  {http://arxiv.org/abs/1510.03386} {arXiv:1510.03386 [hep-th]} \BibitemShut
  {NoStop}%
\bibitem [{\citenamefont {{McCarthy}}(1975)}]{McCarthy1975bms}%
  \BibitemOpen
  \bibfield  {author} {\bibinfo {author} {\bibfnamefont {P.~J.}\ \bibnamefont
  {{McCarthy}}},\ }\href {\doibase 10.1098/rspa.1975.0083} {\bibfield
  {journal} {\bibinfo  {journal} {Proc. Roy. Soc. Lond. A}\ }\textbf {\bibinfo
  {volume} {343}},\ \bibinfo {pages} {489} (\bibinfo {year}
  {1975})}\BibitemShut {NoStop}%
\bibitem [{\citenamefont {Barnich}\ and\ \citenamefont
  {Oblak}(2015)}]{Barnich:2015uva}%
  \BibitemOpen
  \bibfield  {author} {\bibinfo {author} {\bibfnamefont {G.}~\bibnamefont
  {Barnich}}\ and\ \bibinfo {author} {\bibfnamefont {B.}~\bibnamefont
  {Oblak}},\ }\href {\doibase 10.1007/JHEP03(2015)033} {\bibfield  {journal}
  {\bibinfo  {journal} {JHEP}\ }\textbf {\bibinfo {volume} {03}},\ \bibinfo
  {pages} {033} (\bibinfo {year} {2015})},\ \Eprint
  {http://arxiv.org/abs/1502.00010} {arXiv:1502.00010 [hep-th]} \BibitemShut
  {NoStop}%
\bibitem [{\citenamefont {{Strominger}}(2014)}]{Strominger2014bms}%
  \BibitemOpen
  \bibfield  {author} {\bibinfo {author} {\bibfnamefont {A.}~\bibnamefont
  {{Strominger}}},\ }\href {\doibase 10.1007/JHEP07(2014)152} {\bibfield
  {journal} {\bibinfo  {journal} {JHEP}\ }\textbf {\bibinfo {volume} {07}},\
  \bibinfo {eid} {152} (\bibinfo {year} {2014})},\ \Eprint
  {http://arxiv.org/abs/1312.2229} {arXiv:1312.2229 [hep-th]} \BibitemShut
  {NoStop}%
\bibitem [{\citenamefont {Compère}\ \emph {et~al.}(2018)\citenamefont
  {Compère}, \citenamefont {Fiorucci},\ and\ \citenamefont
  {Ruzziconi}}]{Compere:2018ylh}%
  \BibitemOpen
  \bibfield  {author} {\bibinfo {author} {\bibfnamefont {G.}~\bibnamefont
  {Compère}}, \bibinfo {author} {\bibfnamefont {A.}~\bibnamefont {Fiorucci}},
  \ and\ \bibinfo {author} {\bibfnamefont {R.}~\bibnamefont {Ruzziconi}},\
  }\href {\doibase 10.1007/JHEP11(2018)200} {\bibfield  {journal} {\bibinfo
  {journal} {JHEP}\ }\textbf {\bibinfo {volume} {11}},\ \bibinfo {pages} {200}
  (\bibinfo {year} {2018})},\ \bibinfo {note} {[Erratum: JHEP 04, 172
  (2020)]},\ \Eprint {http://arxiv.org/abs/1810.00377} {arXiv:1810.00377
  [hep-th]} \BibitemShut {NoStop}%
\bibitem [{\citenamefont {Comp\`ere}\ \emph {et~al.}(2020)\citenamefont
  {Comp\`ere}, \citenamefont {Oliveri},\ and\ \citenamefont
  {Seraj}}]{Compere:2019gft}%
  \BibitemOpen
  \bibfield  {author} {\bibinfo {author} {\bibfnamefont {G.}~\bibnamefont
  {Comp\`ere}}, \bibinfo {author} {\bibfnamefont {R.}~\bibnamefont {Oliveri}},
  \ and\ \bibinfo {author} {\bibfnamefont {A.}~\bibnamefont {Seraj}},\ }\href
  {\doibase 10.1007/JHEP10(2020)116} {\bibfield  {journal} {\bibinfo  {journal}
  {JHEP}\ }\textbf {\bibinfo {volume} {20}},\ \bibinfo {pages} {116} (\bibinfo
  {year} {2020})},\ \Eprint {http://arxiv.org/abs/1912.03164} {arXiv:1912.03164
  [gr-qc]} \BibitemShut {NoStop}%
\bibitem [{\citenamefont {Newman}\ and\ \citenamefont
  {Penrose}(1962)}]{Newman:1961qr}%
  \BibitemOpen
  \bibfield  {author} {\bibinfo {author} {\bibfnamefont {E.}~\bibnamefont
  {Newman}}\ and\ \bibinfo {author} {\bibfnamefont {R.}~\bibnamefont
  {Penrose}},\ }\href {\doibase 10.1063/1.1724257} {\bibfield  {journal}
  {\bibinfo  {journal} {J. Math. Phys.}\ }\textbf {\bibinfo {volume} {3}},\
  \bibinfo {pages} {566} (\bibinfo {year} {1962})},\ \bibinfo {note} {[Errata:
  J.Math.Phys.4,no.7,998(1963)]}\BibitemShut {NoStop}%
\bibitem [{Note3()}]{Note3}%
  \BibitemOpen
  \bibinfo {note} {$\protect \bar \Psi _2$ is the complex conjugate to $\Psi
  _2$.}\BibitemShut {Stop}%
\bibitem [{\citenamefont {Bieri}\ and\ \citenamefont
  {Garfinkle}(2014)}]{Bieri:2013ada}%
  \BibitemOpen
  \bibfield  {author} {\bibinfo {author} {\bibfnamefont {L.}~\bibnamefont
  {Bieri}}\ and\ \bibinfo {author} {\bibfnamefont {D.}~\bibnamefont
  {Garfinkle}},\ }\href {\doibase 10.1103/PhysRevD.89.084039} {\bibfield
  {journal} {\bibinfo  {journal} {Phys. Rev. D}\ }\textbf {\bibinfo {volume}
  {89}},\ \bibinfo {pages} {084039} (\bibinfo {year} {2014})},\ \Eprint
  {http://arxiv.org/abs/1312.6871} {arXiv:1312.6871 [gr-qc]} \BibitemShut
  {NoStop}%
\bibitem [{\citenamefont {Flanagan}\ \emph {et~al.}(2019)\citenamefont
  {Flanagan}, \citenamefont {Grant}, \citenamefont {Harte},\ and\ \citenamefont
  {Nichols}}]{Flanagan:2018yzh}%
  \BibitemOpen
  \bibfield  {author} {\bibinfo {author} {\bibfnamefont {E.~E.}\ \bibnamefont
  {Flanagan}}, \bibinfo {author} {\bibfnamefont {A.~M.}\ \bibnamefont {Grant}},
  \bibinfo {author} {\bibfnamefont {A.~I.}\ \bibnamefont {Harte}}, \ and\
  \bibinfo {author} {\bibfnamefont {D.~A.}\ \bibnamefont {Nichols}},\ }\href
  {\doibase 10.1103/PhysRevD.99.084044} {\bibfield  {journal} {\bibinfo
  {journal} {Phys. Rev. D}\ }\textbf {\bibinfo {volume} {99}},\ \bibinfo
  {pages} {084044} (\bibinfo {year} {2019})},\ \Eprint
  {http://arxiv.org/abs/1901.00021} {arXiv:1901.00021 [gr-qc]} \BibitemShut
  {NoStop}%
\bibitem [{\citenamefont {Blumenhagen}\ and\ \citenamefont
  {Plauschinn}(2009)}]{Blumenhagen:2009zz}%
  \BibitemOpen
  \bibfield  {author} {\bibinfo {author} {\bibfnamefont {R.}~\bibnamefont
  {Blumenhagen}}\ and\ \bibinfo {author} {\bibfnamefont {E.}~\bibnamefont
  {Plauschinn}},\ }\href {\doibase 10.1007/978-3-642-00450-6} {\bibfield
  {journal} {\bibinfo  {journal} {Lect. Notes Phys.}\ }\textbf {\bibinfo
  {volume} {779}},\ \bibinfo {pages} {1} (\bibinfo {year} {2009})}\BibitemShut
  {NoStop}%
\bibitem [{\citenamefont {Ashtekar}\ and\ \citenamefont
  {Winicour}(1982)}]{Ashtekar1982lk}%
  \BibitemOpen
  \bibfield  {author} {\bibinfo {author} {\bibfnamefont {A.}~\bibnamefont
  {Ashtekar}}\ and\ \bibinfo {author} {\bibfnamefont {J.}~\bibnamefont
  {Winicour}},\ }\href {\doibase 10.1063/1.525283} {\bibfield  {journal}
  {\bibinfo  {journal} {J. Math. Phys.}\ }\textbf {\bibinfo {volume} {23}},\
  \bibinfo {pages} {2410} (\bibinfo {year} {1982})}\BibitemShut {NoStop}%
\end{thebibliography}%

\end{document}